\documentclass[prd,aps,preprintnumbers,floats,floatfix,superscriptaddress,preprintnumbers,showpacs,eqsecnum,nofootinbib,notitlepage]{revtex4-1}

\usepackage[utf8]{inputenc}
\usepackage{latexsym,array,theorem,mathrsfs,bm,float}
\usepackage{psfrag}
\usepackage{amsfonts,amsmath,amssymb,latexsym,array,afterpage,
theorem,mathrsfs,bm,float,epsfig,color,graphicx,tabularx,here,multirow,braket}
\usepackage[colorlinks=true,linkcolor=magenta,citecolor=green]{hyperref}

\usepackage{comment}

\usepackage{ytableau}
\ytableausetup
{mathmode, aligntableaux = center}

\newcommand{\nn}{\nonumber \\}
\newcommand{\p}{\partial}
\newcommand{\pder}[2][]{\frac{\partial#1}{\partial#2}}

\begin{document}
\baselineskip=12pt

\preprint{KUNS-2894, YITP21-115, IPMU21-0063}
\title{Shift-symmetric $SO(N)$ multi-Galileon}
\author{Katsuki \sc{Aoki}}
\email{katsuki.aoki@yukawa.kyoto-u.ac.jp}
\affiliation{Center for Gravitational Physics, Yukawa Institute for Theoretical Physics, Kyoto University, 606-8502, Kyoto, Japan}

\author{Yusuke \sc{Manita}}
\email{manita@tap.scphys.kyoto-u.ac.jp}
\affiliation{Department of physics, Kyoto University, 606-8502, Kyoto, Japan}

\author{Shinji \sc{Mukohyama}}
\email{shinji.mukohyama@yukawa.kyoto-u.ac.jp}
\affiliation{Center for Gravitational Physics, Yukawa Institute for Theoretical Physics, Kyoto University, 606-8502, Kyoto, Japan}
\affiliation{Kavli Institute for the Physics and Mathematics of the Universe (WPI), The University of Tokyo, 277-8583, Chiba, Japan}

\date{\today}

\begin{abstract}
A Poincar\`{e} invariant, local scalar field theory in which the Lagrangian and the equation of motion contain only up to second-order derivatives of the fields is called generalized Galileon. The covariant version of it in four dimensions is called Horndeski theory, and has been vigorously studied in applications to inflation and dark energy. In this paper, we study a class of multi-field extensions of the generalized Galileon theory. By imposing shift and $SO(N)$ symmetries on all the currently known multi-Galileon terms in general dimensions, we find that the structure of the Lagrangian is uniquely determined and parameterized by a series of coupling constants. We also study tensor perturbation in the shift-symmetric $SO(3)$ multi-Galileon theory in four dimensions. The tensor perturbations can obtain a mass term stemming from the same symmetry breaking pattern as the solid inflation. We also find that the shift-symmetric $SO(3)$ multi-Galileon theory gives rise to new cubic interactions of the tensor modes, suggesting the existence of a new type of tensor primordial non-Gaussianity. 
\end{abstract}

\maketitle

\section{Introduction}

The discovery of the late time accelerating expansion of the universe motivates us to modify the gravitational law at cosmological distances while it is necessary to recover general relativity (GR) at shorter distances to pass the solar system experiments of gravity. In this respect, higher-derivative scalar field theories have many interesting properties. For example, in the context of gravity theories with a non-minimally coupled scalar field, even with sufficient modifications at long distances, the nonlinearity of the higher derivative terms can screen the fifth force mediated by the scalar field at short distances~\cite{Nicolis_2004}. This is called the Vainshtein mechanism in analogy with a phenomenon originally found in massive gravity~\cite{VAINSHTEIN1972393}.

It has been known that the systems with higher derivative equations of motion are generically unstable due to an extra degree of freedom, often called Ostrogradsky ghost~\cite{Ostrogradsky:1850fid,woodard2015theorem}, leading to a naive expectation that the Lagrangian should contain at most  first-order derivatives of the field. On the other hand, the Galileon theory introduced by Nicolis et al.~\cite{Nicolis_2009} is a scalar field theory satisfying the symmetry of $\pi\to\pi+b_\mu x^\mu+c$, in which the equations of motion are second-order differential equations, even though the Lagrangian involves the second-order derivative of the field. The Lagrangian is given by \cite{Deffayet_2011}
\begin{align}
  \mathcal{L}_{n+2}^{\mathrm{Gal}}=- a_{n+2}X\delta^{\mu_1\mu_2\cdots\mu_{n}}_{\nu_1\nu_2\cdots\nu_{n}}\p_{\mu_1}\p^{\nu_1}\pi\p_{\mu_2}\p^{\nu_2}\pi\cdots\p_{\mu_n}\p^{\nu_n}\pi~,\label{intro:galileon}
\end{align}
for an integer $n$ satisfying $n<d$,
where $a_{n+2}$ are the arbitrary constants, $d$ is the spacetime dimensions, and $X:=\p_\mu\pi\p^\mu\pi$, respectively. 
In Ref.~\cite{Deffayet_2011}, it was shown that the equations of motion are still second-order differential equations even when $a_{n+2}X$ in \eqref{intro:galileon} is replaced with an arbitrary function $G_{n+2}(\pi,X)$. The resultant theory is called the generalized Galileon theory. In four dimension, the covariantization, i.e.~inclusion of gravity, leads to the Horndeski theory, the most general scalar-tensor theory whose equations of motion contain up to second-order derivatives of the fields~\cite{Deffayet_2011,Horndeski:1974wa,Kobayashi:2011nu}. Hence, in the single field case, the most general theory is obtained through the approach to covariantize the generalized Galileon theory and the Horndeski theory has played a key role for further investigations of the ghost-free higher-order scalar-tensor theories~\cite{Gleyzes:2014dya,Zumalacarregui:2013pma,Langlois:2015cwa,Motohashi:2016ftl,BenAchour:2016fzp,DeFelice:2018ewo}.

The multi-Galileon theory is a multi-scalar field theory that is invariant under the multi-Galilean transformation in the field space, $\pi^i\to\pi^i+b^i_{\mu}x^\mu+c^i$, and in which the equation of motions are second-order differential equations. The Lagrangian is given by \cite{Hinterbichler_2010,Padilla_2011,Padilla_2010}
\begin{align}
  \mathcal{L}^\mathrm{multi-Gal}_{n+2}=-a_{i_1i_2\cdots i_{n+2}}X^{i_{n+1}i_{n+2}}\delta^{\mu_1\mu_2\cdots\mu_{n}}_{\nu_1\nu_2\cdots\nu_{n}}\p_{\mu_1}\p^{\nu_1}\pi^{i_1}\p_{\mu_2}\p^{\nu_2}\pi^{i_2}\cdots\p_{\mu_n}\p^{\nu_n}\pi^{i_n}~,
\end{align}
where $X^{ij}:=\partial_{\mu}\pi^i \partial^{\mu}\pi^j$ and $a_{i_1i_2\cdots i_{n+2}}$ are arbitrary constants that may be regarded as components of a constant tensor in the internal space. The general higher order multi-scalar field theory was expected to be obtained by replacing $a_{i_1i_2\cdots i_{n+2}}X^{i_{n+1}j_{n+2}}$ with an arbitrary tensor made of $\pi^i$ and $X^{ij}$~\cite{Sivanesan_2014}, but a counterexample was also found~\cite{Allys_2017,Kobayashi_2013}. Even now, the most general multi-scalar field theory in which the Lagrangian and the equations of motion contain only up to second-order derivatives of the fields is not known. A covariant version of a class of generalized multi-Galileon theories which was found so far was studied in Ref.~\cite{Padilla_2013,Akama_2017}.

One example of the Galileon theory is derived from the DGP model~\cite{Dvali_2000} as a low energy effective theory on a brane~\cite{Luty:2003vm} in the decoupling limit~\cite{Nicolis_2004}. In this viewpoint, the Poincar\`{e} symmetry of the five dimensional bulk is reduced to the relativistic Galileon symmetry in the effective single-field theory in which the Lagrangian is given by \eqref{intro:galileon} with the replacement of $X$ with $(1\pm X)^{(1-n)/2}$~\cite{Rham_2010}. In the non-relativistic limit $X\ll 1$ it reduces to the original Galileon theory, while the extreme relativistic limit $X\gg 1$ leads to the cuscuta-Galileon~\cite{Rham_2010,Rham_2017}. In the case of a co-dimension $N$ brane, on the other hand, multiple scalar fields $\pi^i$ ($i=1,2,\cdots,N$) are induced on the brane. The Poincar\`{e} symmetry in the bulk recasts as invariance under the relativistic multi-Galileon transformation, $\pi^i\to b_\mu^i x^\mu+c^i+\omega^i{}_j\pi^i\pm b_\mu^j\pi_j\p^\mu\pi^j$, of the scalar fields $\pi^i$, where $\omega^i{}_j$ is the element of $SO(N)$, giving rise to a new internal symmetry that was not present in the single-field Galileon theory~\cite{Hinterbichler_2010}. Then, the internal $SO(N)$ symmetry and the shift symmetry emerge naturally from the higher co-dimension brane in the resultant effective multi-field theory. In the present paper, we shall investigate the properties of a higher derivative multi-scalar theory by imposing the internal shift and $SO(N)$ symmetries.

The shift symmetry and the $SO(N)$ invariance are also well-motivated in phenomenology. In cosmology, a scalar field is often supposed to have a time dependent expectation value to provide a non-trivial dynamics of the universe. The uniform slice of the scalar determines the preferred time slice of the universe, and thus the time diffeomorphism invariance is spontaneously broken by the expectation value of the scalar. Single-scalar theories coupled to gravity with this symmetry breaking pattern are universally described by the effective field theory of inflation~\cite{Cheung_2008}. On the other hand, multiple scalar fields with the shift and $SO(N)$ symmetries allow to have a different symmetry breaking pattern being compatible with homogeneity and isotropy of an ($N+1$)-dimensional universe, called the solid inflation~\cite{Endlich_2013}, where the spatial diffeomorphism invariance is spontaneously broken by spatially dependent expectation values of the scalars. The shift-symmetric $SO(N)$ multi-Galileon theory may serve as a general framework for this symmetry breaking pattern and would provide new phenomenological signatures which are not present in the conventional inflationary models with the broken time diffeomorphism.

The content of this paper is as follows. In section~\ref{sec:rev}, we first review the multi-Galileon theory in the flat spacetime and then formulate the generalized theory coupled to gravity in general $d$-dimensions. In section~\ref{sec:SO(N)Galileon}, we impose the internal shift and $SO(N)$ symmetries to find the concrete form of the Lagrangian. Despite the fact that the theory is composed of multiple scalars, the internal symmetries fix the structure of the generalized Galileon at each order of the fields. We then discuss the tensor perturbations in the shift-symmetric $SO(N)$ Galileon in section~\ref{sec:tensor}, showing that the multi-Galileon terms indeed provide new cubic self-interactions of the tensor modes. Finally, we summarize our results in section~\ref{conclusion}.

\section{Generalized Multi-Galileon Theory}
\label{sec:rev}
\subsection{Flat spacetime}

In this subsection, we consider $N$-scalar field theories in $d$-dimensional Minkowski spacetime of which Lagrangian includes up to second-order derivatives of the scalar fields. The scalar fields are denoted by $\pi^i(x)$ $(i=1,2,\cdots, N)$. Hereafter, $i,j,k,l,\cdots$ are used to denote the labels of the scalars. When the Lagrangian contains the second-order derivatives, its equations of motion are generally fourth-order in derivatives. According to the Ostrogradsky theorem, such a higher derivative system tends to suffer from a ghost instability. However, there are special cases in which the equations of motion are still second-order in derivatives even though the Lagrangian involves the second-order derivatives of $\pi^i$, and we call such a theory {\it generalized multi-Galileon}. The generalized multi-Galileon theory is defined by the following three conditions:
\begin{enumerate}
 \item \label{cond1} the Lagrangian contains up to second-order derivatives of $\pi^i(x)$;
 \item the Lagrangian is a polynomial of the second derivative of $\pi^i(x)$; and
 \item \label{cond3} the field equations contain up to second-order derivatives of $\pi^i(x)$.
\end{enumerate}

The most general Lagrangian of the generalized multi-Galileon theory has not been identified so far, in contrast to the single scalar case. The currently known Lagrangian of the generalized multi-Galileon theory consists of three parts~\cite{Sivanesan_2014,Allys_2017},
\begin{align}
\mathcal{L}=K(\pi^i, \partial \pi^i)+\mathcal{L}_{\rm sin}(\pi^i, \partial \pi^i,\partial^2 \pi^i)+\mathcal{L}_{\rm ext}(\pi^i, \partial \pi^i,\partial^2 \pi^i)
\,,
\label{g_Gal}
\end{align}
which we shall call {\it multi-k-essence}, {\it single-like Galileon}, and {\it extended Galileon}, respectively. The multi-k-essence term contains up to first-order derivative of $\pi^i$ and the latter two terms involve the second-order derivatives explicitly. Thanks to the Lorentz invariance, the multi-k-essence term is represented by an arbitrary function of $\pi^i$ and $X^{ij}:=\partial_\mu\pi^i\partial^\mu\pi^j$. The multi-k-essence term is denoted by $K(\pi^i, \partial \pi^i)=K(\pi^{i},X^{ij})$.

The single-like Galileon terms have forms analogous to those of the single-field generalized Galileon~\cite{Sivanesan_2014},
\begin{align}
  \mathcal{L}_\mathrm{sin}= - \sum_{q=1}^{d-1}G_{k_1\cdots k_q}(\pi^{i}, X^{ij})\delta^{\mu_1\cdots\mu_q}_{\nu_1\cdots\nu_q}\partial^{\nu_1}\partial_{\mu_1}\pi^{k_1} \cdots \partial^{\nu_q}\partial_{\mu_q}\pi^{k_q}~,\label{multig}
\end{align}
where $G_{k_1\cdots k_q}(\pi^{i}, X^{ij})$ are arbitrary functions of $\pi^i$ and $X^{ij}$, and $\delta^{\mu_1\cdots\mu_q}_{\nu_1\cdots\nu_q}=q!\delta^{[\mu_1}_{\nu_1} \cdots \delta^{\mu_q]}_{\nu_q}$. In order that the field equations are second-order in derivatives, the functions $G_{k_1\cdots k_q}(\pi^{i}, X^{ij})$ have to satisfy
\begin{equation}
  G_{k_1\cdots k_q,lm}(\pi^{i}, X^{ij})=G_{(k_1\cdots k_q,lm)}(\pi^{i}, X^{ij})~,
  \label{TaishokaCon}
\end{equation}
where the symbol ``$ {}_{,ij}$'' denotes the derivative with respect to $ X^{ij}$, i.e.
\begin{align}
  G_{,ij}:=\pder[G]{ X^{ij}}=\frac{1}{2}\left(\pder[G]{ X^{ij}}+\pder[G]{ X^{ji}}\right)
  .
\end{align}

The extended Galileon terms do not have any counterpart in the single field theory. In four dimensions, ref.~\cite{Allys_2017} proposed the following terms
\begin{align}
  \mathcal{L}_{\mathrm{ext1}}&=-G_{i_1i_2j_1j_2k_1}(\pi^{i}, X^{ij})\delta^{\mu_1\mu_2\mu_3}_{\nu_1\nu_2\nu_3}\partial_{\mu_1}\pi^{i_1} \partial_{\mu_2}\pi^{i_2}\partial^{\nu_1}\pi^{j_1}\partial^{\nu_2}\pi^{j_2} \partial_{\mu_3}\partial^{\nu_3}\pi^{k_1},\label{ext1}\\
  \mathcal{L}_{\mathrm{ext2}}&=-G_{i_1i_2j_1j_2k_1k_2}(\pi^{i}, X^{ij})\delta^{\mu_1\mu_2\mu_3\mu_4}_{\nu_1\nu_2\nu_3\nu_4}\partial_{\mu_1}\pi^{i_1}\partial_{\mu_2}\pi^{i_2}\partial^{\nu_1}\pi^{j_1}\partial^{\nu_2}\pi^{j_2}\partial_{\mu_3}\partial^{\nu_3}\pi^{k_1}\partial_{\mu_4}\partial^{\nu_4}\pi^{k_2} ,\label{ext2}\\
  \mathcal{L}_{\mathrm{ext3}}&=-G_{i_1i_2i_3j_1j_2j_3k_1}(\pi^{i}, X^{ij})\delta^{\mu_1\mu_2\mu_3\mu_4}_{\nu_1\nu_2\nu_3\nu_4}\partial_{\mu_1}\pi^{i_1}\partial_{\mu_2}\pi^{i_2}\partial_{\mu_3}\pi^{i_3}\partial^{\nu_1}\pi^{j_1}\partial^{\nu_2}\pi^{j_2}\partial^{\nu_3}\pi^{j_3}\partial_{\mu_4}\partial^{\nu_4}\pi^{k_1},\label{ext3}
\end{align}
where $G_{i_1i_2j_1j_2k_1}(\pi^{i},X^{ij}),~G_{i_1i_2j_1j_2k_1k_2}(\pi^{i},X^{ij}),$ and $G_{i_1i_2i_3j_1j_2j_3k_1}(\pi^{i},X^{ij})$ are arbitrary functions of $\pi^i$ and $X^{ij}$. Without loss of generality, we can suppose that these functions satisfy 
\begin{align}
G_{i_1i_2j_1j_2k_1}&= G_{[i_1i_2][j_1j_2]k_1}\,, \quad G_{i_1[i_2j_1j_2]k_1}=0
\,, \\
G_{i_1i_2j_1j_2k_1k_2}&= G_{[i_1i_2][j_1j_2](k_1k_2)}\,, \quad G_{i_1[i_2j_1j_2]k_1k_2}=0
\,, \\
G_{i_1i_2i_3j_1j_2j_3k_1}&= G_{[i_1i_2i_3][j_1j_2j_3]k_1}\,, \quad G_{i_1i_2[i_3j_1j_2j_3]k_1}=0
\,.
\end{align}
The conditions that the field equations are second-order lead to
\begin{align}
G_{i_1i_2j_1j_2k_1,lm}&=G_{i_1i_2j_1j_2(k_1,lm)}
\,, \\
G_{i_1i_2j_1j_2k_1k_2,lm}&= G_{i_1i_2j_1j_2(k_1k_2,lm)}
\,, \\
G_{i_1i_2i_3j_1j_2j_3k_1,lm}&=G_{i_1i_2i_3j_1j_2j_3(k_1,lm)}
\,.
\end{align}
These terms can be generalized straightforwardly to general dimensions as mentioned in~\cite{Allys_2017}. In $d$-dimensions, the extended Galileon terms are given by
\begin{align}
\mathcal{L}_{\rm ext} &=-\sum_{p=2}^{d-1} \sum_{q=1}^{d-p} \mathcal{L}_{(p,q)}
\,, \label{extg} \\
\mathcal{L}_{(p,q)} &=G_{i_1\cdots i_p j_1 \cdots j_p k_1 \cdots k_q} (\pi^i,X^{ij})\mathcal{L}^{i_1\cdots i_p j_1 \cdots j_p k_1 \cdots k_q}
\,,
\label{Lpq_explicit}
\end{align}
with
\begin{align}
\mathcal{L}^{i_1\cdots i_p j_1 \cdots j_p k_1 \cdots k_q}
&:=\delta^{\mu_1\cdots \mu_{p+q}}_{\nu_1 \cdots \nu_{p+q} } 
\partial_{\mu_1}\pi^{i_1}\partial^{\nu_1}\pi^{j_1} \cdots \partial_{\mu_p}\pi^{i_p} \partial^{\nu_p} \pi^{j_p} 
\partial_{\mu_{p+1}}\partial^{\nu_{p+1}}\pi^{k_1} \cdots \partial_{\mu_{p+q}}\partial^{\nu_{p+q}}\pi^{k_q}
\nn
&\,
=
\sum_{\sigma \in S_{p+q}}\Bigg[\mathrm{sgn}(\sigma)\prod_{a=1}^p \p_{\mu_{ \sigma (a)}}\pi^{i_{a}}\p^{\mu_{a}}\pi^{j_{a}}\prod_{b=1}^q \p_{\mu_{ \sigma (p+b)}}\partial^{\mu_{p+b}}\pi^{k_b}\Bigg]
,
\end{align}
where $S_{p+q}$ is the symmetric group of order $p+q$. Since $\mathcal{L}^{i_1\cdots i_p j_1 \cdots j_p k_1 \cdots k_q}$ satisfies
\begin{align}
\mathcal{L}^{i_1\cdots i_p i_1 \cdots i_p  k_1 \cdots k_q} =\mathcal{L}^{[i_1\cdots i_p] [i_1 \cdots i_p]  (k_1 \cdots k_q)}
\,, \quad 
\mathcal{L}^{[i_1\cdots i_p i_1] i_{2} \cdots i_p   k_1 \cdots k_q} =0 
\,,
\label{sym_ext_L}
\end{align}
as shown in Appendix \ref{appendix:second_deriv},
we can suppose
\begin{align}
G_{i_1\cdots i_p j_1 \cdots j_p  k_1 \cdots k_q} =G_{[i_1\cdots i_p] [j_1 \cdots j_p]  (k_1 \cdots k_q)}
\,, \quad
G_{i_1\cdots i_{p-1}[i_p j_1 \cdots j_p]  k_1 \cdots k_q} =0 
\,,
\label{sym_ext}
\end{align}
without loss of generality. The field equations are second-order differential equations if
\begin{align}
G_{i_1\cdots i_p j_1 \cdots j_p  k_1 \cdots k_q, lm}= G_{i_1\cdots i_p j_1 \cdots j_p  (k_1 \cdots k_q, lm)}
\,,
\label{con_ext}
\end{align}
is imposed. The extended terms exist only when $N\geq q$ due to the antisymmetric property of the indices \eqref{sym_ext}.

The extended terms start at $p=2$. One may consider the $p=0,1$ extended terms, but these are not independent from the single-like Galileon terms.
The $p=0$ terms are clearly the same as the single-like Galileon terms and the $p=1$ terms can be shown to be equivalent to the $p=0$ terms up to total derivatives~\cite{Sivanesan_2014}. Each term in \eqref{multig} and \eqref{extg} is specified by $(p,q)$, where $p$ is the number of the antisymmetric indices and $q$ is the number of the power of $\partial^2 \pi$. Therefore, we call it the $(p,q)$ term when we refer to a specific term of the multi-Galileon.

We introduce the collective indices $I_p=[i_1\cdots i_p],~J_p=[j_1 \cdots j_p]$, and $K_q=(k_1 \cdots k_q)$. Here, $p$ indices are antisymmetrized in $I_p$ and $J_p$ while $q$ indices are symmetrized in $K_q$, respectively. The arbitrary function of each generalized Galileon term is simply denoted by $G_{i_1\cdots i_p j_1 \cdots j_p  k_1 \cdots k_q}=G_{I_p J_p K_q}$ where the single-like Galileon is included in $p=0$ (or $p=1$). The $(p,q)$ term \eqref{Lpq_explicit} can be written as
\begin{align}
\mathcal{L}_{(p,q)}=G_{I_p J_p K_q} \mathcal{F}^{I_p J_p} \mathcal{S}^{K_q}
\,, \label{Lpq}
\end{align}
where
\begin{align}
  \mathcal{F}^{I_pJ_p}&:= \partial_{\mu_{1}}\pi^{i_{1}}\partial^{\nu_{1}}\pi^{j_{1}}\cdots \partial_{\mu_{p}}\pi^{i_{p}}\partial^{\nu_{p}}\pi^{j_{p}}\,,
  \label{def_FIJ}
  \\
    \mathcal{S}^{K_{q}}&:= \partial_{\mu_{p+1}}\partial^{\nu_{p+1}}\pi^{k_{1}}\cdots \partial_{\mu_{p+q}}\partial^{\nu_{p+q}}\pi^{k_{q}}\,,
    \label{def_SK}
\end{align}
are the polynomials of the first derivative of $\pi^i$ of degree $2p$ and those of the second derivatives of degree $q$, respectively. 
For simplicity of the notion, we have dropped the spacetime indices in the left-hand sides of \eqref{def_FIJ} and \eqref{def_SK} which are contracted by the generalized Kronecker delta in \eqref{Lpq}.

The functions $G_{k_1 \cdots k_q}$ are completely symmetric in their indices, meaning that the number of independent components is
\begin{align}
 \frac{N(N+1)\cdots (N+q-1)}{q!}
 \label{s_num}
\end{align}
for each $G_{k_1 \cdots k_q}$ before imposing \eqref{TaishokaCon}. On the other hand, the symmetry \eqref{sym_ext} is represented by the following Young tableau,
\begin{align}
\begin{ytableau}
i_1 & j_1 \\
i_2 & j_2 \\
\none[\vdots]  & \none[\vdots] \\
i_p & j_p
\end{ytableau}
\otimes 
\begin{ytableau}
k_1 & k_2 &\none[\cdots] & k_q
\end{ytableau}
\,.
\label{A_young}
\end{align}
The number of independent components of $G_{i_1\cdots i_p j_1 \cdots j_p  k_1 \cdots k_q} $ is thus
\begin{align}
\frac{N(N-1)\cdots (N+1-p) \times (N+1) N \cdots (N+2-p)}{(p+1)! \times  p!}
\times  \frac{N(N+1)\cdots (N+q-1)}{q!}
\label{e_num}
\end{align}
before imposing \eqref{con_ext}. Each $(p,q)$ term consists of a lot of possible arbitrary functions, in general.

We close this subsection by showing that the function $G_{I_p J_p K_q}$ admits the integral under the conditions \eqref{TaishokaCon} and \eqref{con_ext}.
For a given $q \geq 2$, the condition \eqref{con_ext} yields
\begin{align}
G_{I_p J_p k_1\cdots k_{q-2} k_{q-1} k_q, lm}=G_{I_p J_p k_1\cdots k_{q-2} l m , k_{q-1} k_q}
\label{int_con}
\end{align}
as a necessary condition. This condition is nothing but the integrability condition of $G_{I_p J_p K_q}$ in terms of the indices $K_q$. In fact, we can multiply a symmetric $ X_{ij}$-independent quantity $C^{I_p J_p k_1 \cdots k_{q-2} }$ to rewrite \eqref{int_con} as 
\begin{align}
d[(C\cdot G)_{k_{q-1} k_q }d X^{k_{q-1} k_q }]&=0
\end{align}
where
\begin{align}
(C\cdot G)_{k_{q-1} k_q }=C^{I_p J_p k_1 \cdots k_{q-2} } G_{I_r J_r k_1 \cdots k_{q-2} k_{q-1} k_q}
\,,
\end{align}
and $\pi^i$ are understood as constants under the exterior derivative $d$. The Poincar\'{e} lemma states that a closed form is locally an exact form. 
Hence, the Poincar\'{e} lemma guarantees the existence of the integral $G_{I_p J_p k_1 \cdots k_{q-2}}$ such that
\begin{align}
(C\cdot G)_{k_{q-1} k_q }=\frac{\partial }{\partial  X^{k_{q-1} k_q } } (C\cdot G )
\,,
\end{align}
where $(C\cdot G)=C^{I_p J_p k_1 \cdots k_{q-2} }G_{I_p J_p k_1 \cdots k_{q-2}}$, 
at least locally in the sense of the field space.
Since $C^{I_p J_p k_1 \cdots k_{q-2} }$ is arbitrary and independent of $ X^{ij}$, we obtain
\begin{align}
G_{I_p J_p k_1 \cdots k_q}=\frac{\partial}{\partial  X^{k_{q-1} k_{q}}}G_{I_p J_p k_1 \cdots k_{q-2} }
\,.
\label{Ak_to_Ak-1}
\end{align}
The condition \eqref{con_ext} then leads to
\begin{align}
G_{I_p J_p k_1 \cdots k_{q-2}, k_{q-1}k_q} =  G_{I_p J_p (k_1 \cdots k_{q-2}, k_{q-1} k_q) }
\end{align}
for the integral $G_{I_p J_p k_1 \cdots k_{q-2}  } $ as a necessary condition. 
We can thus repeat this procedure and conclude that $G_{I_p J_p K_q}$ is given by 
\begin{align}
G_{I_p J_p K_q }(\pi^{i}, X^{ij})&=\frac{\partial}{\partial  X^{k_1k_2}} \frac{\partial}{\partial  X^{k_3k_4}} \cdots \frac{\partial}{\partial  X^{k_{q-1}k_{q}}} G^{(p,q)}_{I_p J_p} (\pi^{i},X^{ij})
\,,
&(\mathrm{even~}q)
\,,
\label{even_deriv}
\\
G_{I_p J_p K_q}(\pi^{i}, X^{ij})&=\frac{\partial}{\partial  X^{k_2k_3}} \frac{\partial}{\partial  X^{k_4 k_5}} \cdots \frac{\partial}{\partial  X^{k_{q-1}k_q}} G^{(p,q)}_{I_p J_p k_1}(\pi^{i},X^{ij})
\,,
&(\mathrm{odd~}q)
\,,
\end{align}
implying that the function $G_{I_p J_p K_q }$ is generated by differentiating the function $G^{(p,q)}_{I_p J_p}$ or $G^{(p,q)}_{I_p J_p k_1}$ with a smaller number of indices than $G_{I_p J_p K_q }$. The superscripts $(p,q)$ distinguish the generating functions. The existence of the integrals already reduces the number of independent components of the arbitrary functions from \eqref{s_num} and \eqref{e_num} since the index structure of $K_q$ is fixed. We will find further restrictions of the functions by imposing  internal symmetries of the scalars, the shift symmetry and the $SO(N)$ symmetry, in Section \ref{sec:SO(N)Galileon}.

\subsection{Curved spacetime}
We then covariantize the generalized multi-Galileon theory, keeping the conditions \ref{cond1}-\ref{cond3}. If we simply replace the partial derivatives in the flat multi-Galileon Lagrangian by the covariant derivatives, the equations of motion will contain the higher-order derivatives due to the non-commutativity of the covariant derivative. Therefore, it is necessary to introduce counter term(s) to cancel the higher-order derivative in the equation of motion, as is well-known in the Galileon theory of the single field~\cite{Deffayet_2011}.

Following Ref.~\cite{Padilla_2013}, we find that each ghost-free term of the covariant generalized multi-Galileon is given by
\begin{align}
\mathcal{L}_{(p,q)}= \sum_{r=0}^{\lfloor \frac{q}{2}\rfloor} \mathcal{C}_{q,r} \mathcal{L}_{(p,q-2r,r)}
\,,
\label{Lpq_curved}
\end{align}
where $\lfloor \cdots \rfloor$ is the floor function, and
\begin{align}
\mathcal{C}_{q,r}&:= \left(-\frac{1}{8}\right)^{r}\frac{q!}{(q-2r)!r!}
\label{Cnkp}
\,, \\
\mathcal{L}_{(p,q-2r,r)}&:=  G^{(p,q)}_{I_p J_p K_{q-2r}}(\pi^i, X^{ij}) \mathcal{F}^{I_pJ_p}\mathcal{S}^{K_{q-2r}} \mathcal{R}_{(r)}
\,. \label{Lpqr}
\end{align}
We have adopted the notations
\begin{align}
  \mathcal{F}^{I_p,J_p}&:= \nabla_{\mu_{1}}\pi^{i_{1}}\nabla^{\nu_{1}}\pi^{j_{1}}\cdots \nabla_{\mu_{p}}\pi^{i_{p}}\nabla^{\nu_{p}}\pi^{j_{p}}\,,
  \\
    \mathcal{S}^{K_{q-2r}}&:= \nabla_{\mu_{p+1}}\nabla^{\nu_{p+1}}\pi^{k_{1}}\cdots \nabla_{\mu_{p+q-2r}}\nabla^{\nu_{p+q-2r}}\pi^{k_{q-2r}}\,.
    \\
    \mathcal{R}_{(r)} &:= R^{\nu_{p+q-2r+1} \nu_{p+q-2r+2} } {}_{ \mu_{p+q-2r+1} \mu_{p+q-2r+2} } \cdots R^{\nu_{p+q-1} \nu_{p+q} } {}_{ \mu_{p+q-1} \mu_{p+q} } 
    \,.
\end{align}
with implicit spacetime indices contracted by the generalized Kronecker delta $\delta^{\mu_1 \cdots \mu_{p+q}}_{\nu_1 \cdots \nu_{p+q} }$. The functions $G^{(p,q)}_{I_p J_p K_{q-2r}}(\pi^i, X^{ij})$ are given by
\begin{align}
G^{(p,q)}_{I_p J_p K_{q-2r} }&=\frac{\partial}{\partial  X^{k_1k_2}} \cdots \frac{\partial}{\partial  X^{k_{q-2r-1}k_{q-2r}}} G^{(p,q)}_{I_p J_p} (\pi^{i},X^{ij})
\,,
&(\mathrm{even~}q)
\,,
\\
G^{(p,q)}_{I_p J_p K_{q-2r} }&=\frac{\partial}{\partial  X^{k_2k_3}} \cdots \frac{\partial}{\partial  X^{k_{q-2r-1}k_{q-2r}}} G^{(p,q)}_{I_p J_p k_1 }(\pi^{i},X^{ij})
\,,
&(\mathrm{odd~}q)
\,,
\end{align}
with $G^{(p,q)}_{I_p J_p}$ or $G^{(p,q)}_{I_p J_p k_q}$.
Therefore, the covariant $(p,q)$ term \eqref{Lpq_curved} is determined once the function $G^{(p,q)}_{I_p J_p}$ or $G^{(p,q)}_{I_p J_p k_1}$ is specified. \eqref{Lpq_curved} is schematically written as
\begin{align}
\mathcal{L}_{(p,q)}= \sum_{r=0}^{r_{\rm max}} \mathcal{C}_{q,r} \frac{\partial^{r_{\rm max}-r} G^{(p,q)}_{I_p J_p} }{\partial X ^{r_{\rm max}-r} } (\nabla \pi)^{2p} (\nabla \nabla \pi)^{q-2r} R^r
\,,
\label{Lpq_schem}
\end{align}
with $r_{\rm max}=\lfloor \frac{q}{2}\rfloor$
where the $r=0$ term corresponds to the flat generalized Galileon \eqref{Lpq} with the replacement from the partial derivatives to the covariant derivatives and the $r\neq 0$ terms are the counter terms with the curvature to the power of $r$, respectively.

Let us show that the equations of motion of \eqref{Lpq_curved} are differential equations with up to second-order derivatives of the metric as well as $\pi^i$. The variation of \eqref{Lpqr} with respect to $\pi^i$ yields
\begin{align}
\delta \mathcal{L}_{(p,q-2r,r)}= &
\delta G^{(p,q)}_{I_p J_p K_{q-2r}} \mathcal{F}^{I_pJ_p}\mathcal{S}^{K_{q-2r}} \mathcal{R}_{(r)} + G^{(p,q)}_{I_p J_p K_{q-2r}} \delta \mathcal{F}^{I_pJ_p}\mathcal{S}^{K_{q-2r}} \mathcal{R}_{(r)}+ G^{(p,q)}_{I_p J_p K_{q-2r}} \mathcal{F}^{I_pJ_p} \delta \mathcal{S}^{K_{q-2r}} \mathcal{R}_{(r)}
\end{align}
where each term is given by
\begin{align}
\delta G^{(p,q)}_{I_p J_p K_{q-2r}} \mathcal{F}^{I_pJ_p}\mathcal{S}^{K_{q-2r}} \mathcal{R}_{(r)}  &\sim  2G^{(p,q)}_{I_p J_p K_{q-2r+2} }\mathcal{F}^{I_pJ_p}\mathcal{S}^{K_{q-2r}} \mathcal{R}_{(r)} \nabla_{\lambda} \pi^{k_{q-2r+1}} \nabla^{\lambda} \delta \pi^{k_{q-2r+2} } 
\,, \\
G^{(p,q)}_{I_p J_p K_{q-2r}} \delta \mathcal{F}^{I_pJ_p}\mathcal{S}^{K_{q-2r}} \mathcal{R}_{(r)}  &\sim 2p G^{(p,q)}_{I_p J_p K_{q-2r} }\mathcal{F}^{I_{p-1} J_{p-1}}\mathcal{S}^{K_{q-2r}} \mathcal{R}_{(r)} \nabla^{\nu}\pi^{i_p} \nabla_{\mu } \delta \pi^{j_p}
\,, \\
G^{(p,q)}_{I_p J_p K_{q-2r}} \mathcal{F}^{I_pJ_p} \delta \mathcal{S}^{K_{q-2r}} \mathcal{R}_{(r)} &\sim (q-2r) G^{(p,q)}_{I_p J_p K_{q-2r}} \mathcal{F}^{I_pJ_p}  \mathcal{S}^{K_{q-2r-1}} \mathcal{R}_{(r)} \nabla^{\nu} \nabla_{\mu} \delta \pi^{k_{q-2r} }
\,,
\end{align}
respectively.
Here, an equality symbol $\sim$ is introduced, which holds when terms with up to second-order derivatives and boundary terms are ignored. We recall that the spacetime indices $\mu$ and $\nu$ are contracted by the generalized Kronecker delta. Thanks to the contraction (the antisymmetrization of the $\mu$ and $\nu$ indices), the Bianchi identity $\nabla_{[\mu_1} R^{\nu_2 \nu_3}{}_{\mu_2 \mu_3]}   =0$ and the anticommutator of the covariant derivative yield
\begin{align}
\nabla_{\mu}\mathcal{R}_{(r)} &= 0
\,, \label{der_calR} \\
\nabla_{\mu}\mathcal{F}^{I_p J_p} &=0
\,,
\end{align}
and
\begin{align}
\nabla_{\mu} \mathcal{S}^{K_q} &\sim 0 
\,, \label{der1_calS} \\
\nabla^{\nu} \nabla_{\mu}  \mathcal{S}^{K_q} &\sim
 -\frac{1}{4}q  \mathcal{S}^{K_{q-1}}  \nabla_{\lambda} R^{\nu \nu_q }{}_{\mu \mu_q } \nabla^{\lambda} \pi^{k_q}
\,. \label{der2_calS}
\end{align}
On the other hand, we have used the index $\lambda$ which is not contracted by the generalized Kronecker delta. As for the derivatives $\nabla_{\lambda}$, we have
\begin{align}
\nabla_{\lambda} \mathcal{S}^{K_q} &=q \mathcal{S}^{K_{q-1}} \nabla_{\lambda} \nabla^{\nu} \nabla_{\mu}  \pi^{k_q } \,, \\
\nabla_{\lambda} \mathcal{R}_{(r)} &=r \mathcal{R}_{(r-1)} \nabla_{\lambda} R^{\nu_1 \nu_2}{}_{\mu_1 \mu_2} \,. 
\end{align}
Hence, taking integration by parts, we find
\begin{align}
\delta G^{(p,q)}_{I_p J_p K_{q-2r}} \mathcal{F}^{I_pJ_p}\mathcal{S}^{K_{q-2r}} \mathcal{R}_{(r)}  &\sim -2G^{(p,q)}_{I_p J_p K_{q-2r+2} }\mathcal{F}^{I_pJ_p}\nabla^{\lambda} \mathcal{S}^{K_{q-2r}} \mathcal{R}_{(r)} \nabla_{\lambda} \pi^{k_{q-2r+1}} \delta \pi^{k_{q-2r+2} } 
\nn
&-2G^{(p,q)}_{I_p J_p K_{q-2r+2} }\mathcal{F}^{I_pJ_p} \mathcal{S}^{K_{q-2r}} \nabla^{\lambda} \mathcal{R}_{(r)} \nabla_{\lambda} \pi^{k_{q-2r+1}} \delta \pi^{k_{q-2r+2} } 
\label{var_pi1}
\,, \\
G^{(p,q)}_{I_p J_p K_{q-2r}} \delta \mathcal{F}^{I_pJ_p}\mathcal{S}^{K_{q-2r}} \mathcal{R}_{(r)}  &\sim 0
\,, \\
G^{(p,q)}_{I_p J_p K_{q-2r}} \mathcal{F}^{I_pJ_p} \delta \mathcal{S}^{K_{q-2r}} \mathcal{R}_{(r)} &\sim (q-2r) \nabla^{{\nu}} \nabla_{\mu}  G^{(p,q)}_{I_p J_p K_{q-2r}} \mathcal{F}^{I_pJ_p}  \mathcal{S}^{K_{q-2r-1}}  \mathcal{R}_{(r)} \delta \pi^{k_{q-2r} }
\nn
&-\frac{1}{4} \frac{(q-2r)(q-2r-1)}{r+1} G^{(p,q)}_{I_p J_p K_{q-2r}} \mathcal{F}^{I_pJ_p}  \mathcal{S}^{K_{q-2r-2}} \nabla_{\lambda} \mathcal{R}_{(r+1)} 
\nabla^{\lambda} \pi^{k_{q-2r-1} }  \delta \pi^{k_{q-2r} }
\,.
\label{var_pi3}
\end{align}
The first term of \eqref{var_pi3} is
\begin{align}
&\quad
(q-2r) \nabla^{\nu} \nabla_{\mu}  G^{(p,q)}_{I_p J_p K_{q-2r}} \mathcal{F}^{I_pJ_p}  \mathcal{S}^{K_{q-2r-1}}  \mathcal{R}_{(r)} \delta \pi^{k_{q-2r} }
\nn
&\sim 
2(q-2r) G^{(p,q)}_{I_p J_p K_{q-2r+2}} \nabla^{\lambda} \pi^{k_{q-2r+1} } \nabla^{\nu} \nabla_{\mu} \nabla_{\lambda} \pi^{k_{q-2r+2} }\mathcal{F}^{I_pJ_p}  \mathcal{S}^{K_{q-2r-1}}  \mathcal{R}_{(r)} \delta \pi^{k_{q-2r} }
\nn
& \sim 
2(q-2r) G^{(p,q)}_{I_p J_p K_{q-2r+2}}  \mathcal{F}^{I_pJ_p}  \mathcal{S}^{K_{q-2r-1}} \nabla_{\lambda} \nabla^{\nu} \nabla_{\mu}  \pi^{k_{q-2r+2} }  \mathcal{R}_{(r)} \nabla^{\lambda} \pi^{k_{q-2r+1} } \delta \pi^{k_{q-2r} }
\nn
& \sim 
2 G^{(p,q)}_{I_p J_p K_{q-2r+2}} \mathcal{F}^{I_pJ_p}  \nabla_{\lambda} \mathcal{S}^{K_{q-2r}} \mathcal{R}_{(r)} \nabla^{\lambda} \pi^{k_{q-2r+1} }  \delta \pi^{k_{q-2r+2} }
\label{ddG}
\end{align}
which cancel the first term of \eqref{var_pi1}, where we have used the fact that the $k$ indices are symmetrized. As a result, the variation with respect to $\pi^i$ is given by
\begin{align}
\delta \mathcal{L}_{(p,q-2r,r)}&\sim 
-2G^{(p,q)}_{I_p J_p K_{q-2r+2} }\mathcal{F}^{I_pJ_p} \mathcal{S}^{K_{q-2r}} \nabla_{\lambda} \mathcal{R}_{(r)} \nabla^{\lambda} \pi^{k_{q-2r+1}} \delta \pi^{k_{q-2r+2} } 
\nn
&-\frac{1}{4} \frac{(q-2r)(q-2r-1)}{r+1} G^{(p,q)}_{I_p J_p K_{q-2r}} \mathcal{F}^{I_pJ_p}  \mathcal{S}^{K_{q-2r-2}} \nabla_{\lambda} \mathcal{R}_{(r+1)} \nabla^{\lambda} \pi^{k_{q-2r-1} }  \delta \pi^{k_{q-2r} }
\,,
\label{deltaLpqr}
\end{align}
which includes the third-order derivatives of the metric. In particular, we find
\begin{align}
\delta \mathcal{L}_{(p,q,0)}&\sim 
-\frac{1}{4} q(q-1) G^{(p,q)}_{I_p J_p K_{q}} \mathcal{F}^{I_pJ_p}  \mathcal{S}^{K_{q-2}} \nabla_{\lambda} \mathcal{R}_{(1)} \nabla^{\lambda} \pi^{k_{q-1} }  \delta \pi^{k_{q} }
\,,
\label{deltaLr0}
\end{align}
and
\begin{align}
\delta \mathcal{L}_{(p,q-2r_{\rm max},r_{\rm max})}&\sim 
-2G^{(p,q)}_{I_p J_p K_{q-2r_{\rm max}+2} }\mathcal{F}^{I_pJ_p} \mathcal{S}^{K_{q-2r_{\rm max}}} \nabla_{\lambda} \mathcal{R}_{(r_{\rm max})} \nabla^{\lambda} \pi^{k_{q-2r_{\rm max}+1}} \delta \pi^{k_{q-2r_{\rm max}+2} } 
\,,
\label{deltaLrmax}
\end{align}
for $r=0$ and $r=r_{\rm max}=\lfloor \frac{q}{2}\rfloor$, respectively. We then consider the sum \eqref{Lpq_curved} with the coefficients \eqref{Cnkp} that satisfies
\begin{align}
-\frac{1}{4} \frac{(q-2r)(q-2r-1) }{r+1} \mathcal{C}_{q,r}= 2 \mathcal{C}_{q,r+1}
\,.
\label{C_rec}
\end{align}
Focusing on $r'$ and $r'+1$ terms with $0\leq r' < r_{\rm max}$,
\begin{align}
\mathcal{L}_{(p,q)}= \cdots + \mathcal{C}_{q,r'} \mathcal{L}_{(p,q-2r',r')} + \mathcal{C}_{q,r'+1} \mathcal{L}_{(p,q-2r'-2,r'+1)} +\cdots
\,,
\end{align}
one can see that the higher derivative terms cancel each other by the virtue of the recursion relation \eqref{C_rec} because \eqref{deltaLpqr}, \eqref{deltaLr0}, and \eqref{deltaLrmax} indicate
\begin{align}
\mathcal{C}_{q,r'} \delta \mathcal{L}_{(p,q-2r',r')}  &\sim -\frac{1}{4} \frac{(q-2r')(q-2r'-1)}{r'+1} \mathcal{C}_{q,r'}  G^{(p,q)}_{I_p J_p K_{q-2r'}} \mathcal{F}^{I_pJ_p}  \mathcal{S}^{K_{q-2r'-2}} \nabla_{\lambda} \mathcal{R}_{(r'+1)}
\quad\times\nabla^{\lambda} \pi^{k_{q-2r'-1} }  \delta \pi^{k_{q-2r'} } + \cdots
\,,
\end{align}
and
\begin{align}
\mathcal{C}_{q,r'+1} \delta \mathcal{L}_{(p,q-2r'-2,r'+1)} & \sim -2 \mathcal{C}_{q,r'+1} G^{(p,q)}_{I_p J_p K_{q-2r'} }\mathcal{F}^{I_pJ_p} \mathcal{S}^{K_{q-2r'-2}} \nabla_{\lambda} \mathcal{R}_{(r'+1)} 
\quad\times\nabla^{\lambda} \pi^{k_{q-2r'-1}} \delta \pi^{k_{q-2r'} } +\cdots 
\,.
\end{align}
Therefore, the equations of motion of $\pi^i$ contain at most second-order derivatives of the metric as well as the scalars.

Next, we study the variation with respect to the metric to show that the covariant generalized multi-Galileon does not produce any higher-order equations of motion. Only $\mathcal{S}_{(q-2r)}$ and $\mathcal{R}_{(r)}$ in \eqref{Lpqr} contain the derivative of $g_{\mu\nu}$ and the Lagrangian involves at most second-order derivatives of the fields. Hence, we only need to focus on the variation of \eqref{Lpq_curved}, not the action, yielding
\begin{align}
\delta \mathcal{L}_{(p,q-2r,r)} \sim  G^{(p,q)}_{I_p J_p K_{q-2r}} \mathcal{F}^{I_pJ_p} \delta \mathcal{S}^{K_{q-2r}} \mathcal{R}_{(r)} 
+G^{(p,q)}_{I_p J_p K_{q-2r}} \mathcal{F}^{I_pJ_p}\mathcal{S}^{K_{q-2r}} \delta \mathcal{R}_{(r)} 
\,,
\end{align}
where
\begin{align}
G^{(p,q)}_{I_p J_p K_{q-2r}} \mathcal{F}^{I_pJ_p} \delta \mathcal{S}^{K_{q-2r}} \mathcal{R}_{(r)}  &\sim 
(q-2r) G^{(p,q)}_{I_p J_p K_{q-2r}} \mathcal{F}^{I_pJ_p} \mathcal{S}^{K_{q-2r-1}} \mathcal{R}_{(r)} \delta \nabla^{\nu}\nabla_{\mu} \pi^{k_{q-2r}}
\,, \\
G^{(p,q)}_{I_p J_p K_{q-2r}} \mathcal{F}^{I_pJ_p}\mathcal{S}^{K_{q-2r}} \delta \mathcal{R}_{(r)}  &\sim 
r G^{(p,q)}_{I_p J_p K_{q-2r}} \mathcal{F}^{I_pJ_p}\mathcal{S}^{K_{q-2r}} \mathcal{R}_{(r-1)} \delta R^{\nu_1 \nu_2}{}_{\mu_1 \mu_2}
\,,
\end{align} 
and the variations of the second-derivative and the curvature are
\begin{align}
\delta \nabla^{\nu}\nabla_{\mu} \pi^{k_{q-2r}} &=\delta \Gamma^{\lambda \nu}{}_{\mu} \nabla_{\lambda} \pi^{k_{q-2r} }
\,, \\
\delta R^{\nu_1 \nu_2}{}_{\mu_1 \mu_2} &\sim 2\nabla_{\mu_1}\delta \Gamma^{\nu_1 \nu_2}{}_{\mu_2}
\,.
\label{varR}
\end{align}
Here, $\delta \Gamma^{\lambda \nu}{}_{\mu} =g^{\nu\sigma}\delta \Gamma^{\lambda}{}_{\sigma \mu}$ and $\delta \Gamma^{\lambda}{}_{\sigma \mu}$ is the variation of the connection with respect to the metric.
Note that the $\mu$ and $\nu$ indices are antisymmetrized by the contraction, and thus we have not explicitly antisymmetrized the indices in \eqref{varR}. The variation of the connection is
\begin{align}
\delta \Gamma^{\lambda \nu}{}_{\mu} = \frac{1}{2} ( 
\nabla^{\nu} \delta g^{\lambda}{}_{\mu}+\nabla_{\mu} \delta g^{\lambda \nu}
-\nabla^{\lambda} \delta g^{\nu}{}_{\mu})
\end{align}
where the indices of the metric variation $\delta g_{\alpha \beta}$ are raised by the metric. Thanks to the relations \eqref{der_calR}-\eqref{der1_calS}, the first derivatives of $G^{(p,q)}_{I_p J_p K_{q-2r}} \mathcal{F}^{I_pJ_p}  \mathcal{S}^{K_{q-2r}} \mathcal{R}_{(r)} $ with either $\mu$ or $\nu$ indices do not provide any higher derivative terms, implying that
\begin{align}
\delta \nabla^{\nu}\nabla_{\mu} \pi^{k_{q-2r}} &\sim  \frac{1}{2}\nabla^{\lambda} \pi^{k_{q-2r} } \nabla_{\lambda} \delta g^{\nu}{}_{\mu}
\,, \\
\delta R^{\nu_1 \nu_2}{}_{\mu_1 \mu_2} & \sim -2\nabla_{\mu_1} \nabla^{\nu_1} \delta g^{\nu_2}{}_{\mu_2} 
\,.
\end{align}
We then repeat the similar calculations to the variation with respect to $\pi^i$. Integration by parts yields
\begin{align}
G^{(p,q)}_{I_p J_p K_{q-2r}} \mathcal{F}^{I_pJ_p} \delta \mathcal{S}^{K_{q-2r}} \mathcal{R}_{(r)}  &
\sim 
-\frac{q-2r}{2}G^{(p,q)}_{I_p J_p K_{q-2r}} \mathcal{F}^{I_pJ_p} \nabla_{\lambda } \mathcal{S}^{K_{q-2r-1}} \mathcal{R}_{(r)} \nabla^{\lambda} \pi^{k_{q-2r} }  \delta g^{\nu}{}_{\mu}
\nn
& -\frac{q-2r}{2}G^{(p,q)}_{I_p J_p K_{q-2r}} \mathcal{F}^{I_pJ_p}  \mathcal{S}^{K_{q-2r-1}} \nabla_{\lambda } \mathcal{R}_{(r)} \nabla^{\lambda} \pi^{k_{q-2r} } \delta g^{\nu}{}_{\mu}
\end{align}
and
\begin{align}
G^{(p,q)}_{I_p J_p K_{q-2r}} \mathcal{F}^{I_pJ_p}\mathcal{S}^{K_{q-2r}} \delta \mathcal{R}_{(r)}  &\sim 
-2r \nabla_{\mu_1} \nabla^{\nu_1} G^{(p,q)}_{I_p J_p K_{q-2r}} \mathcal{F}^{I_pJ_p}\mathcal{S}^{K_{q-2r}} \mathcal{R}_{(r-1)} \delta g^{\nu_2}{}_{\mu_2} 
\nn
&-2r G^{(p,q)}_{I_p J_p K_{q-2r}} \mathcal{F}^{I_pJ_p}\nabla_{\mu_1} \nabla^{\nu_1}  \mathcal{S}^{K_{q-2r}} \mathcal{R}_{(r-1)} \delta g^{\nu_2}{}_{\mu_2} 
\nn
& \sim -\frac{4r}{p-2r+1}  G^{(p,q)}_{I_p J_p K_{q-2r+2}} \mathcal{F}^{I_pJ_p} \nabla_{\lambda} \mathcal{S}^{K_{q-2r+1}} \mathcal{R}_{(r-1)} \nabla^{\lambda} \pi^{k_{q-2r+2} }\delta g^{\nu}{}_{\mu}
\nn 
&+ \frac{q-2r}{2} G^{(p,q)}_{I_p J_p K_{q-2r}} \mathcal{F}^{I_pJ_p} \mathcal{S}^{K_{q-2r-1}} \nabla_{\lambda} \mathcal{R}_{(r)} \nabla^{\lambda} \pi^{k_{q-2r}  } \delta g^{\nu}{}_{\mu} 
\,,
\end{align}
where \eqref{der2_calS} and \eqref{ddG} are used. Hence, the variation of \eqref{Lpqr} with respect to the metric is 
\begin{align}
\delta \mathcal{L}_{(p,q-2r,r)} & \sim 
-\frac{q-2r}{2}G^{(p,q)}_{I_p J_p K_{q-2r}} \mathcal{F}^{I_pJ_p} \nabla_{\lambda } \mathcal{S}^{K_{q-2r-1}} \mathcal{R}_{(r)} \nabla^{\lambda} \pi^{k_{q-2r} }  \delta g^{\nu}{}_{\mu}
\nn
& -\frac{4r}{p-2r+1}  G^{(p,q)}_{I_p J_p K_{q-2r+2}} \mathcal{F}^{I_pJ_p} \nabla_{\lambda} \mathcal{S}^{K_{q-2r+1}} \mathcal{R}_{(r-1)} \nabla^{\lambda} \pi^{k_{q-2r+2} }\delta g^{\nu}{}_{\mu}
\,.
\end{align}
One can easily confirm that these higher derivative terms are cancelled by summing over $p$ in \eqref{Lpq_curved} by the virtue of the recursion relation \eqref{C_rec}. Therefore, the equations of motion of \eqref{Lpq_curved} contain at most second-order derivatives of the fields.

The covariant generalized multi-Galileon is thus given by the Lagrangian
\begin{align}
\mathcal{L}=K(\pi^i,X^{ij}) - \sum_{p=1}^{d-1} \sum_{q=1}^{d-p} \mathcal{L}_{(p,q)}
\end{align}
with \eqref{Lpq_curved} in general $d$-dimensions. The single-like Galileon and the Lovelock terms are included in $p=1$ (or $p=0$).

\subsection{Explicit covariant Lagrangian in $4$-dimensions}
In four-dimensional spacetime, the Lagrangian is explicitly given by
\begin{align}
\mathcal{L}=K(\pi^i,X^{ij}) - \mathcal{L}_{(0,1)} - \mathcal{L}_{(0,2)} - \mathcal{L}_{(0,3)} - \mathcal{L}_{(2,1)} - \mathcal{L}_{(2,2)} - \mathcal{L}_{(3,1)}
\end{align}
where
\begin{align}
\mathcal{L}_{(0,1)}&=G_{k_1}^{(0,1)} \Box \pi^{k_1}
\,, \\
\mathcal{L}_{(0,2)}&=\frac{ \partial G^{(0,2)}}{\partial X^{k_1 k_2 }} ( \Box \pi^{k_1} \Box \pi^{k_2} - \nabla_{\alpha}\nabla_{\beta} \pi^{k_1} \nabla^{\alpha} \nabla^{\beta} \pi^{k_2})-\frac{1}{2}G^{(0,2)}R
\,, \\
\mathcal{L}_{(0,3)}&= \frac{\partial G_{k_1}^{(0,3)}}{\partial X^{k_2 k_3}} \delta^{\mu_1\mu_2\mu_3}_{\nu_1\nu_2\nu_3}
\nabla_{\mu_1}\nabla^{\nu_1} \pi^{k_1} \nabla_{\mu_2}\nabla^{\nu_2} \pi^{k_2} \nabla_{\mu_3}\nabla^{\nu_3} \pi^{k_3} -3 G^{(0,3)}_{k_1}G^{\mu}{}_{\nu}\nabla_{\mu}\nabla^{\nu} \pi^{k_1} 
\,, \\
\mathcal{L}_{(2,1)}&=G^{(2,1)}_{i_1 i_2 j_2 j_2 k_1}  \delta^{\mu_1\mu_2\mu_3}_{\nu_1\nu_2\nu_3}
\nabla_{\mu_1} \pi^{i_1} \nabla^{\nu_1} \pi^{j_1} \nabla_{\mu_2} \pi^{i_2} \nabla^{\nu_2} \pi^{j_2} \nabla_{\mu_3}\nabla^{\nu_3} \pi^{k_1}
\,, \\
\mathcal{L}_{(2,2)}&= \frac{\partial G^{(2,2)}_{i_1i_2 j_1 j_2} }{\partial X^{k_1 k_2}} \delta^{\mu_1\mu_2\mu_3 \mu_4 }_{\nu_1\nu_2\nu_3\nu_4 }
\nabla_{\mu_1} \pi^{i_1} \nabla^{\nu_1} \pi^{j_1} \nabla_{\mu_2} \pi^{i_2} \nabla^{\nu_2} \pi^{j_2} \nabla_{\mu_3}\nabla^{\nu_3} \pi^{k_1} \nabla_{\mu_4} \nabla^{\nu_4} \pi^{k_2}
\nn
&- G^{(2,2)}_{i_1i_2 j_1 j_2}  G^{\mu_1\mu_2}{}_{\nu_1 \nu_2}
\nabla_{\mu_1} \pi^{i_1} \nabla^{\nu_1} \pi^{j_1} \nabla_{\mu_2} \pi^{i_2} \nabla^{\nu_2} \pi^{j_2} 
\,, \\
\mathcal{L}_{(3,1)}&= G^{(3,1)}_{i_1 i_2 i_3 j_1j_2j_3 k_1} \delta^{\mu_1\mu_2\mu_3 \mu_4 }_{\nu_1\nu_2\nu_3\nu_4 }
\nabla_{\mu_1} \pi^{i_1} \nabla^{\nu_1} \pi^{j_1} \nabla_{\mu_2} \pi^{i_2} \nabla^{\nu_2} \pi^{j_2} \nabla_{\mu_3} \pi^{i_3} \nabla^{\nu_3} \pi^{i_3} \nabla_{\mu_4} \nabla^{\nu_4} \pi^{k_1}
\,.
\end{align}
This agrees with~\cite{Padilla_2013,Akama_2017} . Here, $G^{\mu_1\mu_2}{}_{\nu_1 \nu_2} := \frac{1}{4}\delta^{\mu_1\mu_2\mu_3\mu_4}_{\nu_1\nu_2\nu_3\nu_4}R^{\nu_3 \nu_4}{}_{\mu_3 \mu_4}$ is the double dual of the Riemann tensor and $G^{\mu}{}_{\nu}=G^{\alpha\mu}{}_{\alpha\nu}$ is the Einstein tensor, respectively.

\section{$SO(N)$ multi-Galileon}
\label{sec:SO(N)Galileon}

In this section, we will investigate how the generalized multi-Galileon Lagrangian \eqref{g_Gal} is restricted under the internal global symmetries\footnote{Precisely speaking, we suppose that the Lagrangian is invariant under \eqref{shift_sym} and \eqref{internal_sym} and do not consider the case that the Lagrangian is quasi-invariant, i.e. the Lagrangian is invariant up to a total derivative.}
\begin{align}
  \pi^i&\to \pi^i+c^i,\quad c^{i}=\mathrm{const.}~, \label{shift_sym}\\
  \pi^i&\to O^i{}_j\pi^j,\quad O^i{}_j\in SO(N)~,
  \label{internal_sym}
\end{align}
in addition to the $d$-dimensional Poincar\`{e} symmetry. We introduce the Kronecker delta $\delta_{ij}={\rm diag}[1,1,\cdots ,1]$ and the Levi-Civita symbol $\epsilon_{i_1i_2\cdots i_N}$ in the internal space. The $SO(N)$ indices are raised and lowered by $\delta^{ij}$ and $\delta_{ij}$.
In what follows we discuss the consequences of \eqref{shift_sym} and \eqref{internal_sym} for each part of \eqref{g_Gal} in order.

\subsection{k-essence terms}

Due to the $SO(N)$ symmetry, all internal indices have to be contracted by the use of $\delta^{ij}$, $\delta_{ij}$ and $\epsilon_{i_1i_2\cdots i_N}$ to form a scalar quantity of the $SO(N)$ group. The shift symmetry demands that $K(\pi^{i},X^{ij})$ should not depend on $\pi^i$ explicitly. Since $X_{ij}$ is symmetric in its indices, contractions with $\epsilon_{i_1i_2\cdots i_N}$ identically vanish. Then, $K=K(X^{ij})$ is an arbitrary function of $[X], [X^2], \dots$ where we use the notation
\begin{align}
[X^n]=X^{i_1}{}_{i_2}X^{i_2}{}_{i_3}\cdots X^{i_n}{}_{i_1}
\,.
\end{align}
Any $[ X^n]$ with $n \geq N+1$ is expressed as a polynomial of $[ X],\cdots,[ X^N]$ according to the Cayley-Hamilton theorem. Therefore, we conclude that $K$ is an arbitrary function of $[ X],\cdots,[ X^N]$,
\begin{align}
K=K([ X],\cdots,[ X^N])
\,,
\end{align}
under the shift symmetry and the $SO(N)$ invariance.

\subsection{Single-like terms}
\label{subsec:Shift-symmetric SO(N) Multi-Galileon}

Let us first consider the $SO(N)$ invariance which requires that $G_{k_1\cdots k_q}=G_{K_q}$ should be a tensor of the $SO(N)$ group. As we have explained, the condition \eqref{TaishokaCon} is a necessary condition to have integrals for $q\geq 2$. The $SO(N)$ invariance concludes that the integrals $G^{(0,q)}$ and $G^{(0,q)}_{k_1}$ are an $SO(N)$ scalar and a vector, respectively.

The condition \eqref{TaishokaCon} states that the first derivative of $G_{K_q}$ with respect to $X^{ij}$ has to have completely symmetric indices and its integrals $G_{K_{q'}}$ with $q'<q$ are also symmetric in all their indices. We now consider the second derivative,
\begin{align}
G_{k_1\cdots k_q ,k_{q+1} k_{q+2}, k_{q+3} k_{q+4} } = \frac{\partial }{\partial X^{k_{q+3} k_{q+4}}} G_{k_1\cdots k_q ,k_{q+1}k_{q+2}} = \frac{\partial }{\partial X^{k_{q+3} k_{q+4}}}  \frac{\partial }{\partial X^{k_{q+2} k_{q+1}}} G_{k_1\cdots k_q} \,.
\end{align}
The condition \eqref{TaishokaCon} and the fact that $G_{k_1\cdots k_q ,k_{q+1} k_{q+2}, k_{q+3} k_{q+4} } $ is the second derivative then imply
\begin{align}
G_{k_1\cdots k_q ,k_{q+1} k_{q+2}, k_{q+3} k_{q+4} } &=G_{(k_1\cdots k_q,k_{q+1}k_{q+2}), k_{q+3} k_{q+4} }
\,,
\label{sym_der1}
\\
G_{k_1\cdots k_q, k_{q+1}k_{q+2}, k_{q+3} k_{q+4} } &= G_{k_1\cdots k_q, k_{q+3}k_{q+4},k_{q+1}k_{q+2} }
\,,
\label{sym_der2}
\\
G_{k_1\cdots k_q,k_{q+1}k_{q+2},k_{q+3} k_{q+4} } &= G_{k_1\cdots k_q,k_{q+1}k_{q+2},k_{q+4}k_{q+3} }
\,.
\label{sym_der3}
\end{align}
For every $q \geq 1$, the conditions \eqref{sym_der1}, \eqref{sym_der2}, and \eqref{sym_der3} conclude
\begin{align}
G_{k_1\cdots k_q,k_{q+1}k_{q+2},k_{q+3}k_{q+4} } =G_{(k_1\cdots k_q,k_{q+1}k_{q+2},k_{q+3}k_{q+4}) }
\,.
\end{align}
Indeed, using these three conditions, we can arbitrarily change the indices, for instance
\begin{align}
G_{k_1\cdots k_q,k_{q+1}k_{q+2},k_{q+3}k_{q+4} } &=G_{k_1\cdots k_q,k_{q+3}k_{q+4},k_{q+1}k_{q+2} } 
\nn
&=G_{k_1\cdots k_{q+3},k_{q}k_{q+4},k_{q+1}k_{q+2} } 
\nn
&=G_{k_1\cdots k_{q+3},k_{q+1}k_{q+2},k_{q}k_{q+4} }\,,
\end{align}
where we have used \eqref{sym_der2}, \eqref{sym_der1}, and then \eqref{sym_der2} in each step. As a result, $G_{k_1 k_2\cdots k_q}$ with any $q\geq 1$ has to be obtained by differentiating a generating function $G^{(0,q)}(\pi^{i},X^{ij})$ or $G^{(0,q)}_{k_1} (\pi^{i},X^{ij})$, where the generating functions are defined by the condition that any $n$-th derivative with respect to $X^{ij}$ is symmetric in all their indices,
\begin{align}
G^{(0,q)}_{,k_1 l_1,\cdots , k_{n} l_n} &=G^{(0,q)}_{,(k_1 l_1,\cdots , k_{n} l_{n})}\,, \nn
G^{(0,q)}_{k_0,k_1 l_1,\cdots , k_{n} l_{n}}&=G^{(0,q)}_{(k_0, k_1 l_1,\cdots , k_n l_n)}
\,.
\label{generating_con}
\end{align}
This property holds for any $q$ with $q\geq 1$ when the condition \eqref{TaishokaCon} is imposed.

We then explicitly construct the generating functions. For simplicity of the notation, we shall omit the superscript $(0,q)$. We can expand the generating functions at least locally in terms of $X^{ij}$,
\begin{align}
G(\pi^{i}, X^{ij})&= \sum_{n=0}^{\infty} G^n(\pi^{i},X^{ij})
\label{expansion1}
\,, \\
G_{k_0}(\pi^{i}, X^{ij})&= \sum_{n=0}^{\infty} G_{k_0}^n(\pi^{i},X^{ij})  
\,,
\label{expansion2}
\end{align}
where $G^n$ and $G^n_{k_0}$ are homogeneous polynomials of $X^{ij}$ with the degree $n$; that is, they satisfy $G^n(\pi^{i}, \lambda X^{ij})=\lambda^n G^n(\pi^{i}, X^{ij})$ and $G_{k_0}^n(\pi^{i}, \lambda X^{ij})=\lambda^n G_{k_0}^n(\pi^{i}, X^{ij})$ for an arbitrary $\lambda$. Due to the $SO(N)$ symmetry, $G^n$ and $G^n_{k_0}$ are given by
\begin{align}
    G^n(\pi^{i},X^{ij})&=S_{k_1l_1\cdots k_n l_n}(\pi) X^{k_1 l_1} \cdots X^{k_n l_n}
    \,, 
    \\
    G_{k_0}^n(\pi^{i},X^{ij})&=S_{k_0 k_1 l_1\cdots k_n l_n}(\pi) X^{k_1 l_1} \cdots X^{k_n l_n}
\end{align}
where the $SO(N)$ tensor $S_{k_1 k_2 \cdots k_m}$ satisfies
\begin{align}
S_{k_1 k_2 \cdots k_m}=S_{(k_1 k_2 \cdots k_m)} 
\,, \label{S-symmetric}
\end{align}
because of the conditions \eqref{generating_con}. One can easily construct $S_{k_1 k_2 \cdots k_m}$ from $\pi^i$ and $\delta_{ij}$ by symmetrizing the indices.

We now impose the shift symmetry. This requires that $S_{k_1 k_2 \cdots k_m}$ is constructed by $\delta_{ij}$ only. Then, for each even $m=2n$, $S_{k_1 l_1 \cdots k_n l_n}$ is uniquely determined by
\begin{align}
\Delta_{k_1 l_1 \cdots k_n l_n} := (2n-1)!! \delta_{(k_1l_1} \cdots \delta_{k_n l_n)} =  \delta_{k_1 l_1} \cdots \delta_{k_n l_n} + {\rm permutations} 
\,,
\end{align}
up to an overall constant factor. On the other hand, $S_{k_0k_1l_1\cdots k_nl_n}$ vanishes identically. We define the polynomials by $\bm{X}^0:=1$ and
\begin{align}
\bm{X}^n
&:= \frac{1}{(2n)!!} \Delta_{k_1 l_1 \cdots k_n l_n} X^{k_1 l_1} \cdots X^{k_n l_n}
\,,
\end{align}
for natural numbers $n$, 
and define their derivatives by
\begin{align}
	\bm{X}_{k_1 l_{1} k_{2} l_{2}\cdots k_{m} l_{m}}^{n-m}&:=\frac{\p }{\p X^{k_{1} l_{1}}}\frac{\p }{\p X^{k_{2} l_{2}}}\cdots \frac{\p }{\p X^{k_{m}l_{m}}}\bm{X}^{n}
	\,,
\end{align}
with $m \leq n$. Under the $SO(N)$ symmetry and the shift symmetry the generating functions are given by
\begin{align}
G(X^{ij})&=\sum_{n=0}^{\infty} a_n \bm{X}^n
\,, 
\label{gen_func}
\\
G_{k_0}(X^{ij})&=0
\,,
\end{align}
where $a_n$ are arbitrary constants. All single-like multi-Galileon terms are generated by differentiating the generating functions. In particular, in four dimensions, the single-like multi-Galileon only admits the terms,
\begin{align}
  \mathcal{L}_{\rm sin}=-\mathcal{L}_{(0,2)}
  =\frac{1}{2}a_0 R + \sum_{n=1}^\infty a_n \left[ \frac{1}{2} \bm{X}^n R - \bm{X}^{n-1}_{kl} \left(\Box\pi^k\Box\pi^l-\nabla_{\alpha}\nabla_{\beta}\pi^k \nabla^{\alpha}\nabla^{\beta}\pi^l\right)\right]
  \,,
  \label{cov_singleG}
\end{align}
under the $SO(N)$ symmetry and the shift symmetry. Each of $2n$-point interactions of $\pi^i$ is uniquely determined up to the overall factors $a_n$ by the $SO(N)$ symmetry and the shift symmetry. 

Recall that the number of independent components of the $SO(N)$ tensor $G_{k_1\cdots k_q}$ is originally \eqref{s_num} and the components are functions of the multi-variables $X^{ij}$. However, after demanding the $SO(N)$ symmetry and the shift symmetry, the function $G_{k_1\cdots k_q}$ is determined by the coefficients of the generating function \eqref{gen_func}. Clearly, there is a one-to-one correspondence between the generating function \eqref{gen_func} and a single variable function $f(x)= \sum_{n=0}^{\infty}a_nx^n$. Although there is apparently a huge number of the possibilities of the single-like Galileon terms due to the existence of the multi-variables, the dimension of the theory space is the same as that of the shift symmetric generalized Galileon theory of the single variable when we truncate the Lagrangian up to a finite order of the fields.

\subsection{Extended terms}
We proceed to consider the extended Galileon terms with general $p$ and $q$. The difference from the previous case is the existence of the first $2p$ indices in $G_{i_1\cdots i_p j_1 \cdots j_p k_1\cdots k_q}=G_{I_p J_p K_q}$.

Similarly to the single-like terms, by the use of the Poincar\'{e} lemma, the condition \eqref{con_ext} guarantees the existence of the integrals $G_{I_p J_p}$ and $G_{I_p J_p k_0}$ such that
every $n$-th derivatives with respect to $X^{kl}$ are symmetric in all their last $2n$ or $2n+1$ indices,
\begin{align}
G_{I_p J_p, k_1l_1, \cdots ,k_n l_{n} } &=G_{I_p J_p, (k_1l_1, \cdots ,k_{n}l_{n}) }
\,, 
\label{ext_der1}
\\
G_{I_p J_p k_0, k_1l_1, \cdots ,k_{n}l_{n} } &=G_{I_p J_p (k_0, k_1l_1, \cdots ,k_{n}l_{n}) }
\,,
\label{ext_der2}
\end{align}
respectively. Here, we omit the superscript $(p,q)$ since the property \eqref{ext_der1} or \eqref{ext_der2} holds for any $p$ and $q$. The symmetries of the first $2p$ indices are represented by the following Young tableau,
\begin{align}
\begin{ytableau}
i_1 & j_1 \\
i_2 & j_2 \\
\none[\vdots]  & \none[\vdots] \\
i_p & j_p
\end{ytableau}
\,.
\label{2k_Young}
\end{align}

We expand the generating functions $G_{I_p J_p }$ and $G_{I_p J_p k_0}$ in terms of $X^{kl}$,
\begin{align}
G_{I_p J_p} &= \sum_{n=0}^{\infty} G_{I_p J_p}^n
\,, \\
G_{I_p J_p k_0} &= \sum_{n=0}^{\infty} G_{I_p J_p k_0}^n
\,,
\end{align}
where $G_{I_p J_p}^n$ and $G_{I_p J_p k_0}^n$ are homogeneous polynomials of $X^{ij}$ with the degree $n$. The conditions \eqref{ext_der1} and \eqref{ext_der2} imply that all $X^{ij}$ must be contracted by using $S_{k_1 \cdots k_m}(\pi^{i})$ that satisfies \eqref{S-symmetric}. Since $S_{k_1\cdots k_m}$ is symmetric in all its indices and the first $2p$ indices of $G_{I_p J_p }$ and $G_{I_p J_p k_0}$ have the symmetry \eqref{2k_Young} which can have at most two symmetric indices, the only possible terms are  
\begin{align}
G^n_{i_1\cdots i_p }{}^{j_1 \cdots j_p} &=T_{i_1\cdots i_p }^{j_1 \cdots j_p} S_{k_1 l_1 \cdots k_n l_n }X^{k_1 l_1} \cdots X^{k_n l_n}
+
T_{[i_1\cdots i_{p-1}}^{[j_1 \cdots j_{p-1}} S^{j_p]}{}_{ i_p]k_1 l_1 \cdots k_n l_n }X^{k_1 l_1} \cdots X^{k_n l_n}
\,,
\\
G^n_{i_1\cdots i_p }{}^{j_1 \cdots j_p}{}_{k_0} &= T_{i_1\cdots i_p }^{j_1 \cdots j_p} S_{k_0 k_1 l_1 \cdots k_n l_n }X^{k_1 l_1} \cdots X^{k_n l_n}
+
T_{[i_1\cdots i_{p-1}}^{[j_1 \cdots j_{p-1}} S^{j_p]}{}_{ i_p]k_0 k_1 l_1 \cdots k_n l_n }X^{k_1 l_1} \cdots X^{k_n l_n}
\,,
\end{align}
where $T_{i_1\cdots i_p }^{j_1 \cdots j_p}(\pi^{i})$ is a $SO(N)$ tensor of which indices have the symmetry represented by \eqref{2k_Young}.
The $SO(N)$ symmetry implies that $T_{i_1\cdots i_p }^{j_1 \cdots j_p}(\pi^{i})$ should be constructed from $\pi^i$, $\delta^{ij}$,  $\delta_{ij}$ and $\epsilon_{i_1\cdots i_N}$. The Young tableau \eqref{2k_Young} then implies that there are only two independent pieces of $T_{i_1\cdots i_p }^{j_1 \cdots j_p}(\pi^{i})$, namely
\begin{align}
\delta_{i_1\cdots i_p }^{j_1 \cdots j_p}\,, \quad
\delta_{[i_1\cdots i_{p-1} }^{[j_1 \cdots j_{p-1}}\pi^{j_p]} \pi_{i_p]}
 \,,
 \label{two_pieces}
\end{align}
up to overall scalar functions of $\pi^i \pi_{i} = \delta_{ij}\pi^i \pi^j$, where $\delta_{i_1\cdots i_p }^{j_1 \cdots j_p}=p! \delta_{[i_1}^{j_1} \cdots \delta_{i_p] }^{j_p}$. For instance, for $2p=N$, there exists the tensor $\epsilon_{i_1\cdots i_p j_1 \cdots j_p}$ which satisfies the desired anti-symmetries of the $i$ indices and the $j$ indices, $\epsilon_{i_1\cdots i_p j_1 \cdots j_p}=\epsilon_{[i_1\cdots i_p] [j_1 \cdots j_p]}$, but this is clearly represented by a different Young tableau and then excluded. One may consider the tensor
\begin{align}
\delta_{i_{1}\cdots i_{p+1}}^{j_{1}\cdots j_{p+1}}\pi^{i_{p+1}}\pi_{j_{p+1}}
\,,
\label{linearly_dep}
\end{align}
that satisfies the (anti-)symmetric properties \eqref{2k_Young}. However, \eqref{linearly_dep} is indeed given by a linear combination of the two terms in \eqref{two_pieces}; explicitly,
\begin{align}
  \delta^{j_{1}\cdots j_{p+1}}_{i_{1}\cdots i_{p+1}}\pi^{i_{p+1}}\pi_{j_{p+1}}=&(p+1)\delta^{j_{1}\cdots j_{p}}_{[i_{1}\cdots i_{p}}\delta^{j_{p+1}}_{i_{p+1}]}\pi^{i_{p+1}}\pi_{j_{p+1}}\notag\\
  =&\delta^{j_{1}\cdots j_{p}}_{i_{1}\cdots i_{p}}\pi^{2}-p^{2}\delta^{[j_{1}\cdots j_{p-1}}_{[i_{1}\cdots i_{p-1}}\delta^{j_{p}]}_{|i_{p+1}|}\delta^{j_{p+1}}_{i_{p}]}\pi^{i_{p+1}}\pi_{j_{p+1}}\notag\\
  =&\delta^{j_{1}\cdots j_{p}}_{i_{1}\cdots i_{p}}\pi^{2}-p^{2}\delta_{[i_1\cdots i_{p-1} }^{[j_1 \cdots j_{p-1}}\pi^{j_p]} \pi_{i_p]}
  \,.
\end{align}

Let us now impose the shift symmetry on top of the $SO(N)$ symmetry. The tensors $S_{k_1 \cdots k_m}$ and $T_{i_1\cdots i_p }^{j_1 \cdots j_p}$ are now uniquely determined by $\Delta_{k_1 \cdots k_p}$ and $\delta_{i_1\cdots i_p }^{j_1 \cdots j_p}$, respectively, up to constant factors. Therefore, the generating functions are given by
\begin{align}
G_{i_1\cdots i_p}{}^{ j_1 \cdots j_p} &= \sum_{n=0}^{\infty} b_n \delta_{i_1\cdots i_p}^{j_1 \cdots j_p} \bm{X}^n +\sum_{n=1}^{\infty} c_n \delta_{[i_1\cdots i_{p-1} }^{[j_1 \cdots j_{p-1}} \bm{X}^n_{i_p]}{}^{j_p]}
\,,
\label{gen_ext} \\
G_{i_1\cdots i_p}{}^{ j_1 \cdots j_p k_0} &= 0
\,,
\end{align}
where $b_n$ and $c_n$ are arbitrary constants. Note that $\bm{X}^0_{ij}=\frac{1}{2}\delta_{ij}$ and thus we have set $c_0=0$ without loss of generality by redefining $b_0$. All functions of the extended terms are obtained by differentiating the generating function $G_{i_1\cdots i_r}{}^{ j_1 \cdots j_r}$ which is specified by a couple of the sequences $b_n$ and $c_n$.

Now let us consider the $p=N$ term. For this purpose it is useful to study the $N=2$ case. The $b_n$ terms and the $c_n$ terms are not independent for the $SO(2)$ scalars. The only non-trivial terms are the $p=2$ terms
\begin{align}
G_{i_1i_2}{}^{j_1j_2}=\sum_{n=1}^{\infty} b_n \delta_{i_1 i_2 }^{j_1j_2} \bm{X}^n +\sum_{n=1}^{\infty} c_n \delta_{[i_1}^{[j_1 } \bm{X}^n_{i_2]}{}^{j_2]} 
\end{align}
for $N=2$. The Young diagram of the shape $(2,2)$ for $SO(2)$ has only one independent component, leading to that the $c_n$ terms must be proportional to $ \delta_{i_1 i_2 }^{j_1j_2}$, which can be explicitly shown by computing
\begin{align}
0 \equiv \delta_{j_3}^{i_3} \delta_{[i_1 i_2 }^{[j_1j_2} \bm{X}^n_{i_3]}{}^{j_3]}  = \frac{1}{9} \delta_{i_1 i_2 }^{j_1j_2} \bm{X}^n_{k}{}^k - \frac{4}{9}  \delta_{[i_1}^{[j_1 } \bm{A}^n_{i_2]}{}^{j_2]}
\,.
\end{align}
Since we have the relation $\Delta_{k_1\cdots k_{m} k_{m+1}k_{m+2}} \delta^{k_{m+1} k_{m+2}} \propto \Delta_{k_1\cdots k_{m} }$, the trace of $\bm{X}^n_i{}^{j}$ is proportional to $\bm{X}^n$. Therefore, all $c_n$ terms are proportional to the $b_n$ terms for the $SO(2)$ scalars. This discussion can be generalized into the $p=N$ term for the $SO(N)$ scalars because the Young diagram with the column lengths $(N,N)$ for $SO(N)$ also has only one independent component.

We note that the generating function can be assumed to have the symmetry \eqref{2k_Young} in its indices, but we do not need to impose the symmetry of the indices because the indices are contracted by $\mathcal{L}^{I_p J_p K_q}$ (see \eqref{sym_ext}). We can practically use the generating function
\begin{align}
G_{i_1\cdots i_p}{}^{ j_1 \cdots j_p} &= \sum_{n=0}^{\infty} b_n \delta_{i_1}^{j_1} \cdots \delta_{i_p}^{j_p} \bm{X}^n +\sum_{n=1}^{\infty} c_n \delta_{i_1}^{j_1} \cdots \delta_{ i_{p-1} }^{j_{p-1}} \bm{X}^n_{i_p}{}^{j_p}
\,,
\end{align}
where we have redefined the constants $b_n$ and $c_n$ from \eqref{gen_ext}. As a result, the $SO(N)$ invariance and the shift symmetry conclude that the ghost-free Lagrangian is given by
\begin{align}
\mathcal{L}_{(p,q)} &= \sum_{r=0}^{\frac{q}{2}} \mathcal{C}_{q,r} \mathcal{L}_{(p,q-2r,r)}
\,, \nn
\mathcal{L}_{(p,q-2r,r)} &= \sum_{n=0}^{\infty} b_n \bm{X}^{n-(q-2r)/2}_{K_{q-2r} }\mathcal{F}^{I_p}{}_{I_p} \mathcal{S}^{K_{q-2r} } \mathcal{R}_{(r)}
+ \sum_{n=0}^{\infty} c_n \bm{X}^{n-(q-2r+2)/2}_{K_{q-2r+2}}  \mathcal{F}^{I_{p-1} }{}_{I_{p-1}} \nabla_{\mu} \pi^{k_1} \nabla^{\nu} \pi^{k_2} \mathcal{S}^{K_{q-2r} } \mathcal{R}_{(r)}
\,,
\end{align}
for each non-negative integer $p$ and even number $q$. In the case of $d=4$, the shift-symmetric $SO(N)$ extended multi-Galileon leaves only the $p=2,q=2$ term which is explicitly given by
\begin{align}
  \mathcal{L}_{\mathrm{ext} 2}&=
  b_0  G^{\mu_1\mu_2}{}_{\nu_1\nu_2} \nabla_{\mu_1} \pi_{i_1} \nabla^{\nu_1} \pi^{i_1} \nabla_{\mu_2} \pi_{i_2} \nabla^{\nu_2} \pi^{i_2}
  \nn
  &+
  \sum_{n=1}^{\infty} b_n \Big[ \bm{X}^n G^{\mu_1\mu_2}{}_{\nu_1\nu_2} \nabla_{\mu_1} \pi_{i_1} \nabla^{\nu_1} \pi^{i_1} \nabla_{\mu_2} \pi_{i_2} \nabla^{\nu_2} \pi^{i_2}
  \nn
  &\qquad \qquad - \bm{X}^{n-1}_{k_1 k_2} \delta_{\nu_{1} \nu_{2} \nu_{3} \nu_{4}}^{\mu_{1} \mu_{2} \mu_{3} \mu_{4}} \nabla_{\mu_{1}} \pi_{i_1} \nabla^{\nu_1} \pi^{i_1} \nabla_{\mu_{2}} \pi_{i_2} \nabla^{\nu_{2}} \pi ^{i_2}  \nabla_{\mu_{3}} \nabla^{\nu_{3}} \pi ^{k_1 } \nabla_{\mu_{4}} \nabla^{\nu_{4}} \pi ^{k_2} 
  \Big]
  \nn
  &+
  \sum_{n=1}^{\infty} c_n \Big[ \bm{X}^n_{k_1 k_2} G^{\mu_1\mu_2}{}_{\nu_1\nu_2} \nabla_{\mu_1} \pi_{i_1} \nabla^{\nu_1} \pi^{i_1} \nabla_{\mu_2} \pi^{k_1} \nabla^{\nu_2} \pi^{k_2}
    \nn
  &\qquad \qquad - \bm{X}^{n-1}_{k_1 k_2 k_3 k_4} \delta_{\nu_{1} \nu_{2} \nu_{3} \nu_{4}}^{\mu_{1} \mu_{2} \mu_{3} \mu_{4}} \nabla_{\mu_{1}} \pi_{i_1} \nabla^{\nu_1} \pi^{i_1} \nabla_{\mu_{2}} \pi^{k_1} \nabla^{\nu_{2}} \pi ^{k_2}  \nabla_{\mu_{3}} \nabla^{\nu_{3}} \pi ^{k_3 } \nabla_{\mu_{4}} \nabla^{\nu_{4}} \pi ^{k_4} 
  \Big]
  \,.
  \label{cov_extG}
\end{align}
As already shown in the previous subsection, the single-like terms are specified by an infinite sequence $a_n$. As a result, the generalized multi-Galileon Lagrangian in four spacetime dimensions is determined by three sequences $a_n,b_n,c_n$ and the multi-k-essence term $K=K([X],\cdots, [X^N])$ under the $SO(N)$ symmetry and the shift symmetry. When the spacetime dimension increases, the number of possible terms also increases because higher $(p,q)$ terms are allowed so long as $p$ does not exceed $N$.

\section{Tensor perturbation}
\label{sec:tensor}

We now briefly discuss phenomenological implications of the generalized multi-Galileon theory. The higher derivative scalar-tensor theories have gained attention in the context of cosmology. The scalar fields are usually supposed to have time-dependent background values, $\pi^i =\pi^i(t)$, and cosmological solutions were investigated in, e.g.,~\cite{Kobayashi_2013,Akama_2017}. On the other hand, the existence of multi-fields allows a space-dependent configuration to be compatible with the symmetry of the FLRW universe, which is called the solid inflation in the context of the inflation~\cite{Endlich_2013}. We study such a space-dependent configuration since it may provide signatures qualitatively different from those of single-field theories.

The inflationary universe with a single scalar field can be universally described by the effective field theory of inflation~\cite{Cheung_2008}. The gradient of the scalar field is assumed to be timelike and then the uniform slice of the scalar determines the preferred hypersurface of the spacetime. In the unitary gauge, where the scalar field is identified with the time, the full spacetime diffeomorphism invariance is spontaneously broken down to the invariance under the spatial diffeomorphisms. 

The solid inflation is an inflationary model described by three scalar fields with the internal shift and $SO(3)$ symmetries, where the scalar fields have space-dependent expectation values. Thanks to the internal symmetries, the constant spacelike gradient of the scalars is compatible with the background homogeneity and isotropy if the configuration of the scalar is invariant under an arbitrary spatial rotation up to an internal rotation. In particular, the scalars are identified with the spatial coordinates in the unitary gauge, $\pi^i=x^i$, and then the spatial diffeomorphism invariance is broken. The Nambu-Goldstone boson associated with this symmetry breaking pattern can be interpreted as a solid's phonon, so the inflationary model based on this configuration of three scalar fields is called the solid inflation~\cite{Endlich_2013}.

In the original paper~\cite{Endlich_2013}, the analysis was restricted to the Lagrangian containing up to the first-order derivatives of $\pi^i$, i.e., the Lagrangian is given by the multi-k-essence term in addition to the Einstein-Hilbert term.  We now add the generalized multi-Galileon terms which can be regarded as higher derivative corrections to the original solid inflation. In principle the contributions from the multi-Galileon can be as large as those from the multi-k essence along the line of G-inflation~\cite{Kobayashi:2011nu,Kobayashi_2013} since the multi-Galileon terms are free of ghosts and enjoy the non-renormalization property.

We adopt the unitary gauge, $\pi^i=x^i$, in which we have
\begin{align}
\nabla_{\mu}\pi^i = \delta^i_{\mu}
\,, \quad X^{ij}=g^{\mu\nu}\delta^i_{\mu} \delta^j_{\nu} = g^{ij}
\,.
\end{align}
Hereinafter, the spatial indices are identified with the $SO(3)$ indices, $i,j,k,\cdots$. Since the symmetry breaking pattern is different from the conventional inflationary scenario, there are various new operators that are not present in the effective field theory of inflation. For instance, the multi-k-essence term yields a Lorentz-violating mass term of the tensor mode
\begin{align}
K(X) \supset \delta_{ij}\delta_{kl} X^{ik}X^{jl} = \delta_{ij}\delta_{kl}  g^{ik} g^{jl}
\,.
\label{tensor_mass}
\end{align}
The generalized Galileon terms provide new derivative interactions. In the single field case with the broken time diffeomorphism, a ghost-free term is
\begin{align}
G^{\mu\nu}\partial_{\mu} \pi \partial_{\nu} \pi = G^{00}
\,,
\end{align}
in the unitary gauge $\pi=t \Rightarrow \partial_{\mu}\pi = \delta^0_{\mu}$. On the other hand, the analogous term in the $SO(3)$ Galileon in four dimensions is given by
\begin{align}
\delta_{ij} G^{\mu\nu}\partial_{\mu} \pi^i \partial_{\nu} \pi^j = \delta_{ij} G^{ij}
\,,
\label{Gij}
\end{align}
which comes from the $n=1$ term of the single-like Galileon \eqref{cov_singleG}. Furthermore, the $n=0$ term of the extended Galileon \eqref{cov_extG} is
\begin{align}
\delta_{ij}\delta_{kl}G^{ikjl}
\,.
\label{Gijkl}
\end{align}
The operators \eqref{tensor_mass}, \eqref{Gij} and \eqref{Gijkl} do not respect the spatial diffeomorphism invariance, yielding unique features of the multi-field theory.

For simplicity, we focus on the tensor perturbations and study the Lagrangian up to cubic order around the Minkowski background in the unitary gauge. Defining $\delta \bm{X}^n:= \frac{1}{(2n)!!}\Delta_{k_1 l_1 \cdots k_n l_n} \delta X^{k_1l_1} \cdots \delta X^{k_n l_n}$ with $\delta X^{ij}:=X^{ij}-\delta^{ij}$, we can reorganize the series as
\begin{align}
\sum_{n=0}^{\infty}a_n \bm{X}^n &= \sum_{n=0}^{\infty} A_n \delta \bm{X}^n
\,, \\
\sum_{n=0}^{\infty}b_n \bm{X}^n &= \sum_{n=0}^{\infty} B_n \delta \bm{X}^n 
\,, \\
\sum_{n=1}^{\infty}c_n \bm{X}^n_{ij} &= \sum_{n=1}^{\infty} C_n \delta \bm{X}^n_{ij}
\,,
\end{align}
where
\begin{align}
\delta \bm{X}^{n-1}_{ij} = \frac{\partial}{\partial X^{ij}} \delta \bm{X}^n = \frac{\partial}{\partial \delta X^{ij}} \delta \bm{X}^n
\,,
\end{align}
and $\delta \bm{X}^n$ with more than two indices are defined, accordingly. The multi-k-essence term $K$ can be regarded as a function of $[\delta X],[\delta X^2],[\delta X^3]$, instead of a function of $[X],[X^2],[X^3]$, in which the Minkowski solution is obtained at $K=0,~K_{[\delta X]}=\partial K/\partial [\delta X]=0$. The spacetime metric for the tensor perturbations is given by
\begin{align}
ds^2=-dt^2+e^{\chi_{ij} } dx^i dx^j = -dt^2+\left(\delta_{ij}+\chi_{ij}+ \frac{1}{2} \chi^k{}_i \chi_{kj} +\cdots \right) dx^i dx^j
\,,
\end{align}
where $\chi_{ij}$ is transverse-traceless and the spatial indices are raised and lowered by $\delta^{ij}$ and $\delta_{ij}$.

The quadratic Lagrangian of the tensor mode is
\begin{align}
\mathcal{L}^{(TT)}=\frac{1}{8}\left[ \mathcal{G}_T \dot{\chi}_{ij}\dot{\chi}^{ij}- \mathcal{F}_T\partial_k \chi_{ij} \partial^k \chi^{ij} -\mathcal{M}_T^2 \chi_{ij}\chi^{ij} \right]
\end{align}
where
\begin{align}
\mathcal{G}_T &= A_0-2A_1-4B_0-4B_1-5C_1 \,, \\
\mathcal{F}_T &= A_0-3A_1 \,, \\
\mathcal{M}_T^2 & = -8 K_{[\delta X^2]}
\,.
\end{align}
The no-ghost, no-gradient instability, and no-tachyon conditions are $\mathcal{G}_T>0$, $\mathcal{F}_T>0$, and $\mathcal{M}_T^2>0$, respectively. From the Lagrangian, the speed and the effective mass of the gravitational wave are $c_T^2=\mathcal{F}_T/\mathcal{G}_T$ and $m_T^2=\mathcal{M}_T^2/\mathcal{G}_T$. As we mentioned, the tensor mode acquires the mass due to the symmetry breaking. The generalized Galileon terms are derivative interactions, so they modify the propagation speed of gravitational waves, in general. However, it is interesting that, when both single-like and extended multi-Galileon terms are added, there is a parameter region that satisfies all stability conditions and $c_T^2=1$. This suggests that the generalized multi-Galileon can evade the constraints on the speed of gravitational waves at the late time universe~\cite{LIGOScientific:2017zic} and may be used as a dark energy model as well as an inflationary model, owing to a richer structure of the interactions.

We then consider the cubic Lagrangian of the tensor mode which is given by
\begin{align}
\mathcal{L}^{(TTT)} &= \frac{A_0-4A_1-5A_2}{4}\left(\chi^{ik}\chi^{jl}-\frac{1}{2}\chi^{ij}\chi^{kl}\right) \partial_k \partial_l \chi_{ij}
+ \left(B_0+C_1+\frac{A_2}{2}\right) \left(\chi^{ik}\chi^{jl}-\chi^{ij}\chi^{kl}\right) \partial_k \partial_l \chi_{ij} 
\nn
&+\frac{-2A_1-2A_2+4B_0-4B_2+C_1-7C_2}{16} \chi^i{}_j \chi^j{}_k \ddot{\chi}^k{}_i
+ \frac{A_1+A_2-8B_0-8C_1}{8} \chi^i{}_j \chi^j{}_k \partial_l \partial^l \chi^k{}_i
\nn
&-(K_{[\delta X^2]}+K_{[\delta X^3]})\chi^i{}_j \chi^j{}_k \chi^k{}_i
\,,
\label{LTTT}
\end{align}
after integration by parts. The first term in \eqref{LTTT} is the same cubic self-interaction as GR and the generalized Galileon of the single field, i.e. the Horndeski theory~\cite{Gao:2011vs}, while the remaining terms are new terms due to the spontaneously broken spatial diffeomorphisms. Using the freedom of the perturbative field redefinition, $\chi_{ij} \to \chi'_{ij}=\chi_{ij} + c \chi_{ik}\chi^k{}_j$ with a constant $c$, one may eliminate the term $\chi^i{}_j \chi^j{}_k \ddot{\chi}^k{}_i $. We still have three terms,
\begin{align}
\left(\chi^{ik}\chi^{jl}-\chi^{ij}\chi^{kl}\right) \partial_k \partial_l \chi_{ij} \,, \quad 
\chi^i{}_j \chi^j{}_k \partial_l \partial^l \chi^k{}_i \,, \quad 
\chi^i{}_j \chi^j{}_k \chi^k{}_i
\,,
\end{align}
which are not present in the single-field models. The non-derivative interaction, $\chi^i{}_j \chi^j{}_k \chi^k{}_i$, arises from the multi-k-essence term whereas the derivative interactions come from the generalized multi-Galileon terms. The existence of such derivative interactions is the unique feature of the higher derivative multi-field scalar theory with the spontaneously broken spatial diffeomorphisms and may give rise to a new type of primordial tensor non-Gaussianity when the model is applied to the inflationary universe. \footnote{Eq. \eqref{LTTT} includes all parity invariant graviton three point interactions up to the second-order derivative which are consistent with the symmetry breaking pattern of solid inflation. In fact, in the context of solid inflation, Ref. \cite{Endlich:2013jia,Cabass:2021iii} obtained these terms by the use of the symmetry, and computed their non-Gaussianities. Recently, Ref. \cite{Cabass:2021fnw} have bootstrapped all non-Gaussianities that satisfy symmetry, locality and unitary for EFT of Inflation and general solid inflation. It includes those arising from the new tensor interactions in \eqref{LTTT}. }

We finally comment on the scalar and vector perturbations. As is known in the solid inflation, the scalar and vector perturbations suffer from the strong coupling problem around the Minkowski background or the de Sitter background. The same problem persists even when the multi-Galileon terms are added. On the other hand, as in the original solid inflation, the strong coupling issue may be avoided around the general FLRW spacetime deviating from the exact Minkowski (or de Sitter) spacetime. Another possibility to avoid the strong coupling is to add an additional scalar field having a time-dependent expectation value as discussed in the context of massive gravity~\cite{Dubovsky:2005dw}. However, this implies that the scalar and vector perturbation dynamics would depend on the models that cure the strong coupling. We thus leave them for a future study.

\section{Conclusions and Discussions}
\label{conclusion}

In this paper, we have formulated the generalized multi-Galileon theory in generic $d$-dimensions by imposing the internal shift and $SO(N)$ symmetries. Extending the previous studies~\cite{Sivanesan_2014,Allys_2017,Kobayashi_2013,Padilla_2013,Akama_2017}, we have found the covariant version of the generalized multi-Galileon, in which the equations of motion contain up to second-order derivatives of the metric as well as the scalars. Each ghost-free term in the Lagrangian is specified by a pair of integers ($p$, $q$) and a generating function, where $p$ is the number of the antisymmetric indices and $q$ is the maximum number of the second derivative of the scalar, respectively. The schematic form of the Lagrangian is written in \eqref{Lpq_schem}. Imposing the internal symmetries, we have shown that the generating function is uniquely determined by the symmetries for each degree of $X^{ij}=\nabla_{\mu} \pi^i \nabla^{\mu} \pi^j$. In particular, in four dimensions, the Lagrangian of the shift-symmetric generalized $SO(N)$ multi-Galileon is determined by infinite sequences, $(a_n, b_n, c_n)$, and the explicit forms of the single-like term and the extended term are given by \eqref{cov_singleG} and \eqref{cov_extG}, respectively.

We have also studied the tensor perturbations in the shift-symmetric $SO(3)$ generalized multi-Galileon theory, assuming the symmetry breaking pattern of the solid inflation~\cite{Endlich_2013}. Since the breaking pattern of the spacetime symmetry is different from that in the effective field theory of inflation~\cite{Cheung_2008}, new operators are found in the perturbed Lagrangian. At the quadratic order, the generalized multi-Galileon terms generically modify the propagation speed of the gravitational waves, but there is a special case where the speed of gravitational waves is equal to the speed of light even if the multi-Galileon terms are present. Hence, the multi-Galileon theory can pass the constraints from the gravitational wave observation and can serve as a late time modification of gravity. In addition, there are two new cubic derivative interactions of the tensor modes which may be used as a smoking gun of the inflationary model in the multi-Galileon theory through future observations of the primordial tensor non-Gaussianity.

It is still an open question to identify the most general multi-field theory satisfying the conditions \ref{cond1}-\ref{cond3} listed at the beginning of Section~\ref{sec:rev}. In the single field case, the most general theory in four dimensions has been known as the Horndeski theory~\cite{Deffayet_2011,Horndeski:1974wa,Kobayashi:2011nu}, while the most general equations of motion of the four-dimensional bi-scalar-tensor theory was found in~\cite{Ohashi:2015fma} although the corresponding Lagrangian is not known. As we have found in this paper, the structure of the generalized multi-Galileon term is quite restricted by the internal symmetries, and towards identification of the most general Lagrangian it could be a good first step to identify the most general theory under the internal symmetries.

\acknowledgments
We would like to thank Tsutomu Kobayashi and Takahiro Tanaka for insightful comments. K.A. and Y.M. acknowledges the xTras package \cite{Nutma_2014} which was used for confirming calculations explicitly. The work of K.A. was supported in part by Grants-in-Aid from the Scientific Research Fund of the Japan Society for the Promotion of Science (No.~19J00895 and No.~20K14468). The work of Y.M. was supported by the establishment of university fellowships. The work of S.M. was supported by Japan Society for the Promotion of Science Grants-in-Aid for Scientific Research No. 17H02890, No. 17H06359, and by World Premier International Research Center Initiative, MEXT, Japan.

\appendix

\section{Identities of the generalized Kronecker delta}
\label{appendix:second_deriv}
The symmetric properties \eqref{sym_ext_L} are the consequences of the following identities of the generalized Kronecker delta:
\begin{align}
\delta^{\mu_1 \cdots \mu_k \mu_{k+1} \cdots \mu_{k+l}}_{\nu_1 \cdots \nu_k \mu_{k+1} \cdots \nu_{k+l}}
&\equiv 
\delta^{[\mu_1 \cdots \mu_k] \mu_{k+1} \cdots \mu_{k+l}}_{\nu_1 \cdots \nu_k \mu_{k+1} \cdots \nu_{k+l}}
\equiv 
\delta^{\mu_1 \cdots \mu_k \mu_{k+1} \cdots \mu_{k+l}}_{[\nu_1 \cdots \nu_k] \mu_{k+1} \cdots \nu_{k+l}}
\,, \label{id1}
\\
\delta^{\mu_1 \cdots \mu_i  \cdots \mu_j \cdots \mu_{k}}_{\nu_1 \cdots \nu_i  \cdots \nu_j \cdots \nu_{k}}
&\equiv
\delta^{\mu_1 \cdots \mu_j  \cdots \mu_i \cdots \mu_{k}}_{\nu_1 \cdots \nu_j  \cdots \nu_i \cdots \nu_{k}}
\,, \label{id2}
\end{align}
and
\begin{align}
\delta^{[\mu_{1}\nu_{1}}{}^{\mu_{2}\cdots\mu_{k}]}_{\nu_{2}\cdots\nu_{k}}{}^{(\mu_{k+1}\nu_{k+1})\cdots(\mu_{k+l}\nu_{k+l})}
\equiv 0
\,, \label{id3}
\end{align}
where the superscripts $\nu_i$ are the raised indices,
\begin{align}
\delta^{\mu_1 \cdots \mu_{i-1}}_{\nu_1 \cdots \nu_{i-1}} {}^{\mu_i \nu_i}{}^{\mu_{i+1} \cdots \mu_k}_{\nu_{i+1} \cdots \nu_k}
&= \delta^{\mu_1 \cdots \mu_{i-1} \mu_i \mu_{i+1} \cdots \mu_k}_{\nu_1 \cdots \nu_{i-1} \nu_i' \nu_{i+1} \cdots \nu_k} \eta^{\nu_i \nu_i'}
\,,
\\
\delta^{\mu_1 \cdots \mu_{i-1} \mu_i \nu_i  \cdots \mu_{k}\nu_{k} }_{ \nu_1 \cdots \nu_{i-1} }
&= \delta^{\mu_1 \cdots \mu_{i-1} \mu_i \cdots \mu_k}_{\nu_1 \cdots \nu_{i-1} \nu_i'  \cdots \nu_k' } \eta^{\nu_i \nu_i'} \cdots \eta^{\nu_k \nu_k'}
\,.
\end{align}
The first two identities \eqref{id1} and \eqref{id2} are obvious from the definition of the generalized Kronecker delta. In the rest of this Appendix, we shall prove the third identity \eqref{id3}.

We call $\delta^{\mu_1 \cdots \mu_k}_{\nu_1 \cdots \nu_k}$ the generalized Kronecker delta of order $2k$. Since the generalized Kronecker delta of order $2(k+l)$ is constructed by the product of $k+l$ Kronecker deltas, we have the identity
\begin{align}
\delta^{[\mu_{1}\nu_{1}}{}^{\mu_{2}\cdots\mu_{k+l}]}_{\nu_{2}\cdots\nu_{k+l}}
  \equiv 0
\,,
 \label{AppB:deltaAntisymm}
\end{align}
by antisymmetrizing $k+l+1$ indices. The antisymmetrization can be performed by multiplying the generalized Kronecker delta of order $2(k+l+1)$,
\begin{align}
 \delta^{[\mu_{1}\nu_{1}}{}^{\mu_{2}\cdots\mu_{k+l}]}_{\nu_{2}\cdots\nu_{k+l}}=
 \frac{1}{(k+l+1)!}\delta^{\mu_{1}\nu_{1}\mu_{2}\cdots\mu_{k+l}}_{\mu_{1}^{\prime}\nu_{1}^{\prime}\mu_{2}^{\prime}\cdots \mu_{k+l}^{\prime}}
 \delta^{\mu_{1}^{\prime}\nu_{1}^{\prime}}{}^{\mu_{2}^{\prime}\cdots\mu_{k+l}^{\prime}}_{\nu_{2}\cdots\nu_{k+l}}
 \,.
  \label{AppBother}
\end{align}
We then decompose the generalized Kronecker delta of order $2(k+l+1)$ into those of order $2(k+1)$ and of order $2l$:
\begin{align}
  \delta^{\mu_{1}\nu_{1}\mu_{2}\cdots\mu_{k+l}}_{\mu_{1}^{\prime}\nu_{1}^{\prime}\mu_{2}^{\prime}\cdots \mu_{k+l}^{\prime}}
&=\frac{(k+l+1)!}{(k+1)!l!}\delta^{\mu_{1}\nu_{1}\mu_{2}\cdots\mu_{k}}_{[\mu_{1}^{\prime}\nu_{1}^{\prime}\mu_{2}^{\prime}\cdots\mu_{k}^{\prime}}\delta^{\mu_{k+1}\cdots\mu_{k+l}}_{\mu_{k+1}^{\prime}\cdots\mu_{k+l}^{\prime}]}
  \nn
  &=\frac{(k+l)!}{(k+1)!l!}\left[(k+1)\delta^{\mu_{1}\nu_{1}\mu_{2}\cdots\mu_{k}}_{[\mu_{1}^{\prime}|\nu_{1}^{\prime}|\mu_{2}^{\prime}\cdots\mu_{k}^{\prime}}\delta^{\mu_{k+1}\cdots\mu_{k+l}}_{\mu_{k+1}^{\prime}\cdots\mu_{k+l}^{\prime}]}+(-1)^{k+l-1}l\delta^{\mu_{1}\nu_{1}\mu_{2}\cdots\mu_{k}}_{[\mu_{1}^{\prime}\mu_{2}^{\prime}\mu_{3}^{\prime}\cdots\mu_{k+1}^{\prime}}\delta^{\mu_{k+1}\cdots\mu_{k+l-1}\mu_{k+l}}_{\mu_{k+2}^{\prime}\cdots\mu_{k+l}^{\prime}]\nu_{1}^{\prime}}\right]
  ,
  \label{AppBformula2}
\end{align}
where in the second line the $\mu_i'$ indices are explicitly antisymmetrized while $\nu_1'$ is extracted from the antisymmetrization. Applying this decomposition into \eqref{AppBother}, we find the identity
\begin{align} 
0 &\equiv (k+l+1) \delta^{[\mu_{1}\nu_{1}}{}^{\mu_{2}\cdots\mu_{k+l}]}_{\nu_{2}\cdots\nu_{k+l}}
\nn
&=\frac{1 }{(k+1)!l!}\left[(k+1)\delta^{\mu_{1}\nu_{1}\mu_{2}\cdots\mu_{k}}_{[\mu_{1}^{\prime}|\nu_{1}^{\prime}|\mu_{2}^{\prime}\cdots\mu_{k}^{\prime}}\delta^{\mu_{k+1}\cdots\mu_{k+l}}_{\mu_{k+1}^{\prime}\cdots\mu_{k+l}^{\prime}]}+(-1)^{k+l-1}l\delta^{\mu_{1}\nu_{1}\mu_{2}\cdots\mu_{k}}_{[\mu_{1}^{\prime}\mu_{2}^{\prime}\mu_{3}^{\prime}\cdots\mu_{k+1}^{\prime}}\delta^{\mu_{k+1}\cdots\mu_{k+l-1}\mu_{k+l}}_{\mu_{k+2}^{\prime}\cdots\mu_{k+l}^{\prime}]\nu_{1}^{\prime}}\right]
\delta^{\mu_{1}^{\prime}\nu_{1}^{\prime}}{}^{\mu_{2}^{\prime}\cdots\mu_{k+l}^{\prime}}_{\nu_{2}\cdots\nu_{k+l}}
\nn
&=
\frac{1 }{(k+1)!l!} \left[ (k+1) \delta^{\mu_{1}\nu_{1}\mu_{2}\cdots\mu_{k}}_{\mu_{1}^{\prime}\nu_{1}^{\prime}\mu_{2}^{\prime}\cdots\mu_{k}^{\prime}}\delta^{\mu_{k+1}\cdots\mu_{k+l}}_{\mu_{k+1}^{\prime}\cdots\mu_{k+l}^{\prime}}
+(-1)^{k+l-1} l
\delta^{\mu_{1}\nu_{1}\mu_{2}\cdots\mu_{k}}_{\mu_{1}^{\prime}\mu_{2}^{\prime}\mu_{3}^{\prime}\cdots\mu_{k+1}^{\prime}}\delta^{\mu_{k+1}\cdots\mu_{k+l-1}\mu_{k+l}}_{\mu_{k+2}^{\prime}\cdots\mu_{k+l}^{\prime}\nu_{1}^{\prime}}\right]
\delta^{\mu_{1}^{\prime}\nu_{1}^{\prime}}{}^{\mu_{2}^{\prime}\cdots\mu_{k+l}^{\prime}}_{\nu_{2}\cdots\nu_{k+l}}
\nn
&=
(k+1) \delta^{[\mu_{1}\nu_{1} } {}^{\mu_{2}\cdots \mu_k] }_{\nu_{2} \cdots \nu_k} {}^{\mu_{k+1} \cdots \mu_{k+1}}_{\nu_{k+1 } \cdots \nu_{k+l} }
+l \delta^{\mu_1\nu_1\mu_2 \cdots \mu_{k}}_{\nu_2 \nu_3 \nu_4\cdots \nu_{k+2}} {}^{ [\mu_{k+1} \cdots \mu_{k+l-2} \mu_{k+l-1} \mu_{k+l} ] }_{\, \nu_{k+3} \cdots \nu_{k+l} }
\,,
\label{antidelta_dec}
\end{align}
that is,
\begin{align}
  \delta^{[\mu_{1}\nu_{1}}{}^{\mu_{2}\cdots\mu_{k}]}_{\nu_{2}\cdots\nu_{n}}{}^{\mu_{k+1}\cdots\mu_{k+l}}_{\nu_{k+1}\cdots\nu_{k+l}}
  \equiv 
  -\frac{l}{k+1}\delta^{\mu_{1}\nu_{1}\mu_{2}\cdots\mu_{k}}_{\nu_{2}\nu_{3}\nu_{4}\cdots\nu_{k+2}}{}^{[\mu_{k+1}\cdots\mu_{k+l-2}}_{\nu_{k+3}\cdots\nu_{k+l}}{}^{\mu_{k+l-1}\mu_{k+l}]}\,,
  \label{gdelta_id}
\end{align}
where from the second line to the third line of \eqref{antidelta_dec} we have used the fact that the $\mu_i'$ indices of $\delta^{\mu_{1}^{\prime}\nu_{1}^{\prime}}{}^{\mu_{2}^{\prime}\cdots\mu_{k+l}^{\prime}}_{\nu_{2}\cdots\nu_{k+l}}$ are the superscripts of the generalized Kronecker delta of order $2(k+l)$ and thus are completely antisymmetric.
We finally symmetrize the pairs $\mu_i$ and $\nu_i$ with $i\geq k+1$ after raising the indices $\nu_i$. Recall that the right-hand side of \eqref{gdelta_id} is
\begin{align}
-\frac{l}{k+1}\delta^{\mu_{1}\nu_{1}\mu_{2}\cdots\mu_{k}}_{\nu_{2}\nu_{3}\nu_{4}\cdots\nu_{k+2}}{}^{[\mu_{k+1}\cdots\mu_{k+l-2}}_{\nu_{k+3}\cdots\nu_{k+l}}{}^{\mu_{k+l-1}\mu_{k+l}]}
=-\frac{l}{k+1}
\delta^{\mu_{1}\nu_{1}\mu_{2}\cdots\mu_{k}}_{\nu_{2}\nu_{3}\nu_{4}\cdots\nu_{k+2}}{}^{[\mu_{k+1}\cdots\mu_{k+l-2}}_{\, \nu_{k+3}\cdots\nu_{k+l}}{}^{\mu_{k+l-1}}_{\rho } \eta^{\mu_{k+l}] \rho }
\,,
\end{align}
meaning that one of the superscripts $\mu_i$ with $i\geq k+1$ in the right-hand side is a raised subscript of the generalized Kronecker delta. The subscripts $\nu_i~(i\geq k+1)$ are also the subscripts of the generalized Kronecker delta and the subscripts of the generalized Kronecker delta are completely antisymmetric.
This implies that the right-hand side of \eqref{gdelta_id} vanishes when all $\mu_i$ with $i\geq k+1$ are symmetrized by pairing with $\nu_i$, and then the left-hand side also vanishes thanks to the identity \eqref{gdelta_id}. As a result, we obtain the identity \eqref{id3} which yields the second one of \eqref{sym_ext_L}.

\bibliography{paperv2}

\begin{thebibliography}{36}%
\makeatletter
\providecommand \@ifxundefined [1]{%
 \@ifx{#1\undefined}
}%
\providecommand \@ifnum [1]{%
 \ifnum #1\expandafter \@firstoftwo
 \else \expandafter \@secondoftwo
 \fi
}%
\providecommand \@ifx [1]{%
 \ifx #1\expandafter \@firstoftwo
 \else \expandafter \@secondoftwo
 \fi
}%
\providecommand \natexlab [1]{#1}%
\providecommand \enquote  [1]{``#1''}%
\providecommand \bibnamefont  [1]{#1}%
\providecommand \bibfnamefont [1]{#1}%
\providecommand \citenamefont [1]{#1}%
\providecommand \href@noop [0]{\@secondoftwo}%
\providecommand \href [0]{\begingroup \@sanitize@url \@href}%
\providecommand \@href[1]{\@@startlink{#1}\@@href}%
\providecommand \@@href[1]{\endgroup#1\@@endlink}%
\providecommand \@sanitize@url [0]{\catcode `\\12\catcode `\$12\catcode
  `\&12\catcode `\#12\catcode `\^12\catcode `\_12\catcode `\%12\relax}%
\providecommand \@@startlink[1]{}%
\providecommand \@@endlink[0]{}%
\providecommand \url  [0]{\begingroup\@sanitize@url \@url }%
\providecommand \@url [1]{\endgroup\@href {#1}{\urlprefix }}%
\providecommand \urlprefix  [0]{URL }%
\providecommand \Eprint [0]{\href }%
\providecommand \doibase [0]{http://dx.doi.org/}%
\providecommand \selectlanguage [0]{\@gobble}%
\providecommand \bibinfo  [0]{\@secondoftwo}%
\providecommand \bibfield  [0]{\@secondoftwo}%
\providecommand \translation [1]{[#1]}%
\providecommand \BibitemOpen [0]{}%
\providecommand \bibitemStop [0]{}%
\providecommand \bibitemNoStop [0]{.\EOS\space}%
\providecommand \EOS [0]{\spacefactor3000\relax}%
\providecommand \BibitemShut  [1]{\csname bibitem#1\endcsname}%
\let\auto@bib@innerbib\@empty
\bibitem [{\citenamefont {Nicolis}\ and\ \citenamefont
  {Rattazzi}(2004)}]{Nicolis_2004}%
  \BibitemOpen
  \bibfield  {author} {\bibinfo {author} {\bibfnamefont {A.}~\bibnamefont
  {Nicolis}}\ and\ \bibinfo {author} {\bibfnamefont {R.}~\bibnamefont
  {Rattazzi}},\ }\href {\doibase 10.1088/1126-6708/2004/06/059} {\bibfield
  {journal} {\bibinfo  {journal} {JHEP}\ }\textbf {\bibinfo {volume} {06}},\
  \bibinfo {pages} {059} (\bibinfo {year} {2004})},\ \Eprint
  {http://arxiv.org/abs/hep-th/0404159} {arXiv:hep-th/0404159} \BibitemShut
  {NoStop}%
\bibitem [{\citenamefont {Vainshtein}(1972)}]{VAINSHTEIN1972393}%
  \BibitemOpen
  \bibfield  {author} {\bibinfo {author} {\bibfnamefont {A.}~\bibnamefont
  {Vainshtein}},\ }\href {\doibase
  https://doi.org/10.1016/0370-2693(72)90147-5} {\bibfield  {journal} {\bibinfo
   {journal} {Physics Letters B}\ }\textbf {\bibinfo {volume} {39}},\ \bibinfo
  {pages} {393 } (\bibinfo {year} {1972})}\BibitemShut {NoStop}%
\bibitem [{\citenamefont {Ostrogradsky}(1850)}]{Ostrogradsky:1850fid}%
  \BibitemOpen
  \bibfield  {author} {\bibinfo {author} {\bibfnamefont {M.}~\bibnamefont
  {Ostrogradsky}},\ }\href@noop {} {\bibfield  {journal} {\bibinfo  {journal}
  {Mem. Acad. St. Petersbourg}\ }\textbf {\bibinfo {volume} {6}},\ \bibinfo
  {pages} {385} (\bibinfo {year} {1850})}\BibitemShut {NoStop}%
\bibitem [{\citenamefont {Woodard}(2015)}]{woodard2015theorem}%
  \BibitemOpen
  \bibfield  {author} {\bibinfo {author} {\bibfnamefont {R.~P.}\ \bibnamefont
  {Woodard}},\ }\href {\doibase 10.4249/scholarpedia.32243} {\bibfield
  {journal} {\bibinfo  {journal} {Scholarpedia}\ }\textbf {\bibinfo {volume}
  {10}},\ \bibinfo {pages} {32243} (\bibinfo {year} {2015})},\ \Eprint
  {http://arxiv.org/abs/1506.02210} {arXiv:1506.02210 [hep-th]} \BibitemShut
  {NoStop}%
\bibitem [{\citenamefont {Nicolis}\ \emph {et~al.}(2009)\citenamefont
  {Nicolis}, \citenamefont {Rattazzi},\ and\ \citenamefont
  {Trincherini}}]{Nicolis_2009}%
  \BibitemOpen
  \bibfield  {author} {\bibinfo {author} {\bibfnamefont {A.}~\bibnamefont
  {Nicolis}}, \bibinfo {author} {\bibfnamefont {R.}~\bibnamefont {Rattazzi}}, \
  and\ \bibinfo {author} {\bibfnamefont {E.}~\bibnamefont {Trincherini}},\
  }\href {\doibase 10.1103/PhysRevD.79.064036} {\bibfield  {journal} {\bibinfo
  {journal} {Phys. Rev. D}\ }\textbf {\bibinfo {volume} {79}},\ \bibinfo
  {pages} {064036} (\bibinfo {year} {2009})},\ \Eprint
  {http://arxiv.org/abs/0811.2197} {arXiv:0811.2197 [hep-th]} \BibitemShut
  {NoStop}%
\bibitem [{\citenamefont {Deffayet}\ \emph {et~al.}(2011)\citenamefont
  {Deffayet}, \citenamefont {Gao}, \citenamefont {Steer},\ and\ \citenamefont
  {Zahariade}}]{Deffayet_2011}%
  \BibitemOpen
  \bibfield  {author} {\bibinfo {author} {\bibfnamefont {C.}~\bibnamefont
  {Deffayet}}, \bibinfo {author} {\bibfnamefont {X.}~\bibnamefont {Gao}},
  \bibinfo {author} {\bibfnamefont {D.}~\bibnamefont {Steer}}, \ and\ \bibinfo
  {author} {\bibfnamefont {G.}~\bibnamefont {Zahariade}},\ }\href {\doibase
  10.1103/PhysRevD.84.064039} {\bibfield  {journal} {\bibinfo  {journal} {Phys.
  Rev. D}\ }\textbf {\bibinfo {volume} {84}},\ \bibinfo {pages} {064039}
  (\bibinfo {year} {2011})},\ \Eprint {http://arxiv.org/abs/1103.3260}
  {arXiv:1103.3260 [hep-th]} \BibitemShut {NoStop}%
\bibitem [{\citenamefont {Horndeski}(1974)}]{Horndeski:1974wa}%
  \BibitemOpen
  \bibfield  {author} {\bibinfo {author} {\bibfnamefont {G.~W.}\ \bibnamefont
  {Horndeski}},\ }\href {\doibase 10.1007/BF01807638} {\bibfield  {journal}
  {\bibinfo  {journal} {Int. J. Theor. Phys.}\ }\textbf {\bibinfo {volume}
  {10}},\ \bibinfo {pages} {363} (\bibinfo {year} {1974})}\BibitemShut
  {NoStop}%
\bibitem [{\citenamefont {Kobayashi}\ \emph {et~al.}(2011)\citenamefont
  {Kobayashi}, \citenamefont {Yamaguchi},\ and\ \citenamefont
  {Yokoyama}}]{Kobayashi:2011nu}%
  \BibitemOpen
  \bibfield  {author} {\bibinfo {author} {\bibfnamefont {T.}~\bibnamefont
  {Kobayashi}}, \bibinfo {author} {\bibfnamefont {M.}~\bibnamefont
  {Yamaguchi}}, \ and\ \bibinfo {author} {\bibfnamefont {J.}~\bibnamefont
  {Yokoyama}},\ }\href {\doibase 10.1143/PTP.126.511} {\bibfield  {journal}
  {\bibinfo  {journal} {Prog. Theor. Phys.}\ }\textbf {\bibinfo {volume}
  {126}},\ \bibinfo {pages} {511} (\bibinfo {year} {2011})},\ \Eprint
  {http://arxiv.org/abs/1105.5723} {arXiv:1105.5723 [hep-th]} \BibitemShut
  {NoStop}%
\bibitem [{\citenamefont {Gleyzes}\ \emph {et~al.}(2015)\citenamefont
  {Gleyzes}, \citenamefont {Langlois}, \citenamefont {Piazza},\ and\
  \citenamefont {Vernizzi}}]{Gleyzes:2014dya}%
  \BibitemOpen
  \bibfield  {author} {\bibinfo {author} {\bibfnamefont {J.}~\bibnamefont
  {Gleyzes}}, \bibinfo {author} {\bibfnamefont {D.}~\bibnamefont {Langlois}},
  \bibinfo {author} {\bibfnamefont {F.}~\bibnamefont {Piazza}}, \ and\ \bibinfo
  {author} {\bibfnamefont {F.}~\bibnamefont {Vernizzi}},\ }\href {\doibase
  10.1103/PhysRevLett.114.211101} {\bibfield  {journal} {\bibinfo  {journal}
  {Phys. Rev. Lett.}\ }\textbf {\bibinfo {volume} {114}},\ \bibinfo {pages}
  {211101} (\bibinfo {year} {2015})},\ \Eprint {http://arxiv.org/abs/1404.6495}
  {arXiv:1404.6495 [hep-th]} \BibitemShut {NoStop}%
\bibitem [{\citenamefont {Zumalac\'arregui}\ and\ \citenamefont
  {Garc\'\i{}a-Bellido}(2014)}]{Zumalacarregui:2013pma}%
  \BibitemOpen
  \bibfield  {author} {\bibinfo {author} {\bibfnamefont {M.}~\bibnamefont
  {Zumalac\'arregui}}\ and\ \bibinfo {author} {\bibfnamefont {J.}~\bibnamefont
  {Garc\'\i{}a-Bellido}},\ }\href {\doibase 10.1103/PhysRevD.89.064046}
  {\bibfield  {journal} {\bibinfo  {journal} {Phys. Rev. D}\ }\textbf {\bibinfo
  {volume} {89}},\ \bibinfo {pages} {064046} (\bibinfo {year} {2014})},\
  \Eprint {http://arxiv.org/abs/1308.4685} {arXiv:1308.4685 [gr-qc]}
  \BibitemShut {NoStop}%
\bibitem [{\citenamefont {Langlois}\ and\ \citenamefont
  {Noui}(2016)}]{Langlois:2015cwa}%
  \BibitemOpen
  \bibfield  {author} {\bibinfo {author} {\bibfnamefont {D.}~\bibnamefont
  {Langlois}}\ and\ \bibinfo {author} {\bibfnamefont {K.}~\bibnamefont
  {Noui}},\ }\href {\doibase 10.1088/1475-7516/2016/02/034} {\bibfield
  {journal} {\bibinfo  {journal} {JCAP}\ }\textbf {\bibinfo {volume} {02}},\
  \bibinfo {pages} {034} (\bibinfo {year} {2016})},\ \Eprint
  {http://arxiv.org/abs/1510.06930} {arXiv:1510.06930 [gr-qc]} \BibitemShut
  {NoStop}%
\bibitem [{\citenamefont {Motohashi}\ \emph {et~al.}(2016)\citenamefont
  {Motohashi}, \citenamefont {Noui}, \citenamefont {Suyama}, \citenamefont
  {Yamaguchi},\ and\ \citenamefont {Langlois}}]{Motohashi:2016ftl}%
  \BibitemOpen
  \bibfield  {author} {\bibinfo {author} {\bibfnamefont {H.}~\bibnamefont
  {Motohashi}}, \bibinfo {author} {\bibfnamefont {K.}~\bibnamefont {Noui}},
  \bibinfo {author} {\bibfnamefont {T.}~\bibnamefont {Suyama}}, \bibinfo
  {author} {\bibfnamefont {M.}~\bibnamefont {Yamaguchi}}, \ and\ \bibinfo
  {author} {\bibfnamefont {D.}~\bibnamefont {Langlois}},\ }\href {\doibase
  10.1088/1475-7516/2016/07/033} {\bibfield  {journal} {\bibinfo  {journal}
  {JCAP}\ }\textbf {\bibinfo {volume} {07}},\ \bibinfo {pages} {033} (\bibinfo
  {year} {2016})},\ \Eprint {http://arxiv.org/abs/1603.09355} {arXiv:1603.09355
  [hep-th]} \BibitemShut {NoStop}%
\bibitem [{\citenamefont {Ben~Achour}\ \emph {et~al.}(2016)\citenamefont
  {Ben~Achour}, \citenamefont {Crisostomi}, \citenamefont {Koyama},
  \citenamefont {Langlois}, \citenamefont {Noui},\ and\ \citenamefont
  {Tasinato}}]{BenAchour:2016fzp}%
  \BibitemOpen
  \bibfield  {author} {\bibinfo {author} {\bibfnamefont {J.}~\bibnamefont
  {Ben~Achour}}, \bibinfo {author} {\bibfnamefont {M.}~\bibnamefont
  {Crisostomi}}, \bibinfo {author} {\bibfnamefont {K.}~\bibnamefont {Koyama}},
  \bibinfo {author} {\bibfnamefont {D.}~\bibnamefont {Langlois}}, \bibinfo
  {author} {\bibfnamefont {K.}~\bibnamefont {Noui}}, \ and\ \bibinfo {author}
  {\bibfnamefont {G.}~\bibnamefont {Tasinato}},\ }\href {\doibase
  10.1007/JHEP12(2016)100} {\bibfield  {journal} {\bibinfo  {journal} {JHEP}\
  }\textbf {\bibinfo {volume} {12}},\ \bibinfo {pages} {100} (\bibinfo {year}
  {2016})},\ \Eprint {http://arxiv.org/abs/1608.08135} {arXiv:1608.08135
  [hep-th]} \BibitemShut {NoStop}%
\bibitem [{\citenamefont {De~Felice}\ \emph {et~al.}(2018)\citenamefont
  {De~Felice}, \citenamefont {Langlois}, \citenamefont {Mukohyama},
  \citenamefont {Noui},\ and\ \citenamefont {Wang}}]{DeFelice:2018ewo}%
  \BibitemOpen
  \bibfield  {author} {\bibinfo {author} {\bibfnamefont {A.}~\bibnamefont
  {De~Felice}}, \bibinfo {author} {\bibfnamefont {D.}~\bibnamefont {Langlois}},
  \bibinfo {author} {\bibfnamefont {S.}~\bibnamefont {Mukohyama}}, \bibinfo
  {author} {\bibfnamefont {K.}~\bibnamefont {Noui}}, \ and\ \bibinfo {author}
  {\bibfnamefont {A.}~\bibnamefont {Wang}},\ }\href {\doibase
  10.1103/PhysRevD.98.084024} {\bibfield  {journal} {\bibinfo  {journal} {Phys.
  Rev. D}\ }\textbf {\bibinfo {volume} {98}},\ \bibinfo {pages} {084024}
  (\bibinfo {year} {2018})},\ \Eprint {http://arxiv.org/abs/1803.06241}
  {arXiv:1803.06241 [hep-th]} \BibitemShut {NoStop}%
\bibitem [{\citenamefont {Hinterbichler}\ \emph {et~al.}(2010)\citenamefont
  {Hinterbichler}, \citenamefont {Trodden},\ and\ \citenamefont
  {Wesley}}]{Hinterbichler_2010}%
  \BibitemOpen
  \bibfield  {author} {\bibinfo {author} {\bibfnamefont {K.}~\bibnamefont
  {Hinterbichler}}, \bibinfo {author} {\bibfnamefont {M.}~\bibnamefont
  {Trodden}}, \ and\ \bibinfo {author} {\bibfnamefont {D.}~\bibnamefont
  {Wesley}},\ }\href {\doibase 10.1103/PhysRevD.82.124018} {\bibfield
  {journal} {\bibinfo  {journal} {Phys. Rev. D}\ }\textbf {\bibinfo {volume}
  {82}},\ \bibinfo {pages} {124018} (\bibinfo {year} {2010})},\ \Eprint
  {http://arxiv.org/abs/1008.1305} {arXiv:1008.1305 [hep-th]} \BibitemShut
  {NoStop}%
\bibitem [{\citenamefont {Padilla}\ \emph {et~al.}(2011)\citenamefont
  {Padilla}, \citenamefont {Saffin},\ and\ \citenamefont
  {Zhou}}]{Padilla_2011}%
  \BibitemOpen
  \bibfield  {author} {\bibinfo {author} {\bibfnamefont {A.}~\bibnamefont
  {Padilla}}, \bibinfo {author} {\bibfnamefont {P.~M.}\ \bibnamefont {Saffin}},
  \ and\ \bibinfo {author} {\bibfnamefont {S.-Y.}\ \bibnamefont {Zhou}},\
  }\href {\doibase 10.1103/PhysRevD.83.045009} {\bibfield  {journal} {\bibinfo
  {journal} {Phys. Rev. D}\ }\textbf {\bibinfo {volume} {83}},\ \bibinfo
  {pages} {045009} (\bibinfo {year} {2011})},\ \Eprint
  {http://arxiv.org/abs/1008.0745} {arXiv:1008.0745 [hep-th]} \BibitemShut
  {NoStop}%
\bibitem [{\citenamefont {Padilla}\ \emph {et~al.}(2010)\citenamefont
  {Padilla}, \citenamefont {Saffin},\ and\ \citenamefont
  {Zhou}}]{Padilla_2010}%
  \BibitemOpen
  \bibfield  {author} {\bibinfo {author} {\bibfnamefont {A.}~\bibnamefont
  {Padilla}}, \bibinfo {author} {\bibfnamefont {P.~M.}\ \bibnamefont {Saffin}},
  \ and\ \bibinfo {author} {\bibfnamefont {S.-Y.}\ \bibnamefont {Zhou}},\
  }\href {\doibase 10.1007/JHEP12(2010)031} {\bibfield  {journal} {\bibinfo
  {journal} {JHEP}\ }\textbf {\bibinfo {volume} {12}},\ \bibinfo {pages} {031}
  (\bibinfo {year} {2010})},\ \Eprint {http://arxiv.org/abs/1007.5424}
  {arXiv:1007.5424 [hep-th]} \BibitemShut {NoStop}%
\bibitem [{\citenamefont {Sivanesan}(2014)}]{Sivanesan_2014}%
  \BibitemOpen
  \bibfield  {author} {\bibinfo {author} {\bibfnamefont {V.}~\bibnamefont
  {Sivanesan}},\ }\href {\doibase 10.1103/PhysRevD.90.104006} {\bibfield
  {journal} {\bibinfo  {journal} {Phys. Rev. D}\ }\textbf {\bibinfo {volume}
  {90}},\ \bibinfo {pages} {104006} (\bibinfo {year} {2014})},\ \Eprint
  {http://arxiv.org/abs/1307.8081} {arXiv:1307.8081 [gr-qc]} \BibitemShut
  {NoStop}%
\bibitem [{\citenamefont {Allys}(2017)}]{Allys_2017}%
  \BibitemOpen
  \bibfield  {author} {\bibinfo {author} {\bibfnamefont {E.}~\bibnamefont
  {Allys}},\ }\href {\doibase 10.1103/PhysRevD.95.064051} {\bibfield  {journal}
  {\bibinfo  {journal} {Phys. Rev. D}\ }\textbf {\bibinfo {volume} {95}},\
  \bibinfo {pages} {064051} (\bibinfo {year} {2017})},\ \Eprint
  {http://arxiv.org/abs/1612.01972} {arXiv:1612.01972 [hep-th]} \BibitemShut
  {NoStop}%
\bibitem [{\citenamefont {Kobayashi}\ \emph {et~al.}(2013)\citenamefont
  {Kobayashi}, \citenamefont {Tanahashi},\ and\ \citenamefont
  {Yamaguchi}}]{Kobayashi_2013}%
  \BibitemOpen
  \bibfield  {author} {\bibinfo {author} {\bibfnamefont {T.}~\bibnamefont
  {Kobayashi}}, \bibinfo {author} {\bibfnamefont {N.}~\bibnamefont
  {Tanahashi}}, \ and\ \bibinfo {author} {\bibfnamefont {M.}~\bibnamefont
  {Yamaguchi}},\ }\href {\doibase 10.1103/PhysRevD.88.083504} {\bibfield
  {journal} {\bibinfo  {journal} {Phys. Rev. D}\ }\textbf {\bibinfo {volume}
  {88}},\ \bibinfo {pages} {083504} (\bibinfo {year} {2013})},\ \Eprint
  {http://arxiv.org/abs/1308.4798} {arXiv:1308.4798 [hep-th]} \BibitemShut
  {NoStop}%
\bibitem [{\citenamefont {Padilla}\ and\ \citenamefont
  {Sivanesan}(2013)}]{Padilla_2013}%
  \BibitemOpen
  \bibfield  {author} {\bibinfo {author} {\bibfnamefont {A.}~\bibnamefont
  {Padilla}}\ and\ \bibinfo {author} {\bibfnamefont {V.}~\bibnamefont
  {Sivanesan}},\ }\href {\doibase 10.1007/JHEP04(2013)032} {\bibfield
  {journal} {\bibinfo  {journal} {JHEP}\ }\textbf {\bibinfo {volume} {04}},\
  \bibinfo {pages} {032} (\bibinfo {year} {2013})},\ \Eprint
  {http://arxiv.org/abs/1210.4026} {arXiv:1210.4026 [gr-qc]} \BibitemShut
  {NoStop}%
\bibitem [{\citenamefont {Akama}\ and\ \citenamefont
  {Kobayashi}(2017)}]{Akama_2017}%
  \BibitemOpen
  \bibfield  {author} {\bibinfo {author} {\bibfnamefont {S.}~\bibnamefont
  {Akama}}\ and\ \bibinfo {author} {\bibfnamefont {T.}~\bibnamefont
  {Kobayashi}},\ }\href {\doibase 10.1103/PhysRevD.95.064011} {\bibfield
  {journal} {\bibinfo  {journal} {Phys. Rev. D}\ }\textbf {\bibinfo {volume}
  {95}},\ \bibinfo {pages} {064011} (\bibinfo {year} {2017})},\ \Eprint
  {http://arxiv.org/abs/1701.02926} {arXiv:1701.02926 [hep-th]} \BibitemShut
  {NoStop}%
\bibitem [{\citenamefont {Dvali}\ \emph {et~al.}(2000)\citenamefont {Dvali},
  \citenamefont {Gabadadze},\ and\ \citenamefont {Porrati}}]{Dvali_2000}%
  \BibitemOpen
  \bibfield  {author} {\bibinfo {author} {\bibfnamefont {G.}~\bibnamefont
  {Dvali}}, \bibinfo {author} {\bibfnamefont {G.}~\bibnamefont {Gabadadze}}, \
  and\ \bibinfo {author} {\bibfnamefont {M.}~\bibnamefont {Porrati}},\ }\href
  {\doibase 10.1016/S0370-2693(00)00669-9} {\bibfield  {journal} {\bibinfo
  {journal} {Phys. Lett. B}\ }\textbf {\bibinfo {volume} {485}},\ \bibinfo
  {pages} {208} (\bibinfo {year} {2000})},\ \Eprint
  {http://arxiv.org/abs/hep-th/0005016} {arXiv:hep-th/0005016} \BibitemShut
  {NoStop}%
\bibitem [{\citenamefont {Luty}\ \emph {et~al.}(2003)\citenamefont {Luty},
  \citenamefont {Porrati},\ and\ \citenamefont {Rattazzi}}]{Luty:2003vm}%
  \BibitemOpen
  \bibfield  {author} {\bibinfo {author} {\bibfnamefont {M.~A.}\ \bibnamefont
  {Luty}}, \bibinfo {author} {\bibfnamefont {M.}~\bibnamefont {Porrati}}, \
  and\ \bibinfo {author} {\bibfnamefont {R.}~\bibnamefont {Rattazzi}},\ }\href
  {\doibase 10.1088/1126-6708/2003/09/029} {\bibfield  {journal} {\bibinfo
  {journal} {JHEP}\ }\textbf {\bibinfo {volume} {09}},\ \bibinfo {pages} {029}
  (\bibinfo {year} {2003})},\ \Eprint {http://arxiv.org/abs/hep-th/0303116}
  {arXiv:hep-th/0303116} \BibitemShut {NoStop}%
\bibitem [{\citenamefont {de~Rham}\ and\ \citenamefont
  {Tolley}(2010)}]{Rham_2010}%
  \BibitemOpen
  \bibfield  {author} {\bibinfo {author} {\bibfnamefont {C.}~\bibnamefont
  {de~Rham}}\ and\ \bibinfo {author} {\bibfnamefont {A.~J.}\ \bibnamefont
  {Tolley}},\ }\href {\doibase 10.1088/1475-7516/2010/05/015} {\bibfield
  {journal} {\bibinfo  {journal} {JCAP}\ }\textbf {\bibinfo {volume} {05}},\
  \bibinfo {pages} {015} (\bibinfo {year} {2010})},\ \Eprint
  {http://arxiv.org/abs/1003.5917} {arXiv:1003.5917 [hep-th]} \BibitemShut
  {NoStop}%
\bibitem [{\citenamefont {de~Rham}\ and\ \citenamefont
  {Motohashi}(2017)}]{Rham_2017}%
  \BibitemOpen
  \bibfield  {author} {\bibinfo {author} {\bibfnamefont {C.}~\bibnamefont
  {de~Rham}}\ and\ \bibinfo {author} {\bibfnamefont {H.}~\bibnamefont
  {Motohashi}},\ }\href {\doibase 10.1103/PhysRevD.95.064008} {\bibfield
  {journal} {\bibinfo  {journal} {Phys. Rev. D}\ }\textbf {\bibinfo {volume}
  {95}},\ \bibinfo {pages} {064008} (\bibinfo {year} {2017})},\ \Eprint
  {http://arxiv.org/abs/1611.05038} {arXiv:1611.05038 [hep-th]} \BibitemShut
  {NoStop}%
\bibitem [{\citenamefont {Cheung}\ \emph {et~al.}(2008)\citenamefont {Cheung},
  \citenamefont {Creminelli}, \citenamefont {Fitzpatrick}, \citenamefont
  {Kaplan},\ and\ \citenamefont {Senatore}}]{Cheung_2008}%
  \BibitemOpen
  \bibfield  {author} {\bibinfo {author} {\bibfnamefont {C.}~\bibnamefont
  {Cheung}}, \bibinfo {author} {\bibfnamefont {P.}~\bibnamefont {Creminelli}},
  \bibinfo {author} {\bibfnamefont {A.}~\bibnamefont {Fitzpatrick}}, \bibinfo
  {author} {\bibfnamefont {J.}~\bibnamefont {Kaplan}}, \ and\ \bibinfo {author}
  {\bibfnamefont {L.}~\bibnamefont {Senatore}},\ }\href {\doibase
  10.1088/1126-6708/2008/03/014} {\bibfield  {journal} {\bibinfo  {journal}
  {JHEP}\ }\textbf {\bibinfo {volume} {03}},\ \bibinfo {pages} {014} (\bibinfo
  {year} {2008})},\ \Eprint {http://arxiv.org/abs/0709.0293} {arXiv:0709.0293
  [hep-th]} \BibitemShut {NoStop}%
\bibitem [{\citenamefont {Endlich}\ \emph {et~al.}(2013)\citenamefont
  {Endlich}, \citenamefont {Nicolis},\ and\ \citenamefont
  {Wang}}]{Endlich_2013}%
  \BibitemOpen
  \bibfield  {author} {\bibinfo {author} {\bibfnamefont {S.}~\bibnamefont
  {Endlich}}, \bibinfo {author} {\bibfnamefont {A.}~\bibnamefont {Nicolis}}, \
  and\ \bibinfo {author} {\bibfnamefont {J.}~\bibnamefont {Wang}},\ }\href
  {\doibase 10.1088/1475-7516/2013/10/011} {\bibfield  {journal} {\bibinfo
  {journal} {JCAP}\ }\textbf {\bibinfo {volume} {10}},\ \bibinfo {pages} {011}
  (\bibinfo {year} {2013})},\ \Eprint {http://arxiv.org/abs/1210.0569}
  {arXiv:1210.0569 [hep-th]} \BibitemShut {NoStop}%
\bibitem [{\citenamefont {Abbott}\ \emph {et~al.}(2017)\citenamefont {Abbott}
  \emph {et~al.}}]{LIGOScientific:2017zic}%
  \BibitemOpen
  \bibfield  {author} {\bibinfo {author} {\bibfnamefont {B.~P.}\ \bibnamefont
  {Abbott}} \emph {et~al.} (\bibinfo {collaboration} {LIGO Scientific, Virgo,
  Fermi-GBM, INTEGRAL}),\ }\href {\doibase 10.3847/2041-8213/aa920c} {\bibfield
   {journal} {\bibinfo  {journal} {Astrophys. J. Lett.}\ }\textbf {\bibinfo
  {volume} {848}},\ \bibinfo {pages} {L13} (\bibinfo {year} {2017})},\ \Eprint
  {http://arxiv.org/abs/1710.05834} {arXiv:1710.05834 [astro-ph.HE]}
  \BibitemShut {NoStop}%
\bibitem [{\citenamefont {Gao}\ \emph {et~al.}(2011)\citenamefont {Gao},
  \citenamefont {Kobayashi}, \citenamefont {Yamaguchi},\ and\ \citenamefont
  {Yokoyama}}]{Gao:2011vs}%
  \BibitemOpen
  \bibfield  {author} {\bibinfo {author} {\bibfnamefont {X.}~\bibnamefont
  {Gao}}, \bibinfo {author} {\bibfnamefont {T.}~\bibnamefont {Kobayashi}},
  \bibinfo {author} {\bibfnamefont {M.}~\bibnamefont {Yamaguchi}}, \ and\
  \bibinfo {author} {\bibfnamefont {J.}~\bibnamefont {Yokoyama}},\ }\href
  {\doibase 10.1103/PhysRevLett.107.211301} {\bibfield  {journal} {\bibinfo
  {journal} {Phys. Rev. Lett.}\ }\textbf {\bibinfo {volume} {107}},\ \bibinfo
  {pages} {211301} (\bibinfo {year} {2011})},\ \Eprint
  {http://arxiv.org/abs/1108.3513} {arXiv:1108.3513 [astro-ph.CO]} \BibitemShut
  {NoStop}%
\bibitem [{\citenamefont {Endlich}\ \emph {et~al.}(2014)\citenamefont
  {Endlich}, \citenamefont {Horn}, \citenamefont {Nicolis},\ and\ \citenamefont
  {Wang}}]{Endlich:2013jia}%
  \BibitemOpen
  \bibfield  {author} {\bibinfo {author} {\bibfnamefont {S.}~\bibnamefont
  {Endlich}}, \bibinfo {author} {\bibfnamefont {B.}~\bibnamefont {Horn}},
  \bibinfo {author} {\bibfnamefont {A.}~\bibnamefont {Nicolis}}, \ and\
  \bibinfo {author} {\bibfnamefont {J.}~\bibnamefont {Wang}},\ }\href {\doibase
  10.1103/PhysRevD.90.063506} {\bibfield  {journal} {\bibinfo  {journal} {Phys.
  Rev. D}\ }\textbf {\bibinfo {volume} {90}},\ \bibinfo {pages} {063506}
  (\bibinfo {year} {2014})},\ \Eprint {http://arxiv.org/abs/1307.8114}
  {arXiv:1307.8114 [hep-th]} \BibitemShut {NoStop}%
\bibitem [{\citenamefont {Cabass}(2021)}]{Cabass:2021iii}%
  \BibitemOpen
  \bibfield  {author} {\bibinfo {author} {\bibfnamefont {G.}~\bibnamefont
  {Cabass}},\ }\href@noop {} {\  (\bibinfo {year} {2021})},\ \Eprint
  {http://arxiv.org/abs/2103.09816} {arXiv:2103.09816 [hep-th]} \BibitemShut
  {NoStop}%
\bibitem [{\citenamefont {Cabass}\ \emph {et~al.}(2021)\citenamefont {Cabass},
  \citenamefont {Pajer}, \citenamefont {Stefanyszyn},\ and\ \citenamefont
  {Supe\l{}}}]{Cabass:2021fnw}%
  \BibitemOpen
  \bibfield  {author} {\bibinfo {author} {\bibfnamefont {G.}~\bibnamefont
  {Cabass}}, \bibinfo {author} {\bibfnamefont {E.}~\bibnamefont {Pajer}},
  \bibinfo {author} {\bibfnamefont {D.}~\bibnamefont {Stefanyszyn}}, \ and\
  \bibinfo {author} {\bibfnamefont {J.}~\bibnamefont {Supe\l{}}},\ }\href@noop
  {} {\  (\bibinfo {year} {2021})},\ \Eprint {http://arxiv.org/abs/2109.10189}
  {arXiv:2109.10189 [hep-th]} \BibitemShut {NoStop}%
\bibitem [{\citenamefont {Dubovsky}\ \emph {et~al.}(2005)\citenamefont
  {Dubovsky}, \citenamefont {Tinyakov},\ and\ \citenamefont
  {Tkachev}}]{Dubovsky:2005dw}%
  \BibitemOpen
  \bibfield  {author} {\bibinfo {author} {\bibfnamefont {S.~L.}\ \bibnamefont
  {Dubovsky}}, \bibinfo {author} {\bibfnamefont {P.~G.}\ \bibnamefont
  {Tinyakov}}, \ and\ \bibinfo {author} {\bibfnamefont {I.~I.}\ \bibnamefont
  {Tkachev}},\ }\href {\doibase 10.1103/PhysRevD.72.084011} {\bibfield
  {journal} {\bibinfo  {journal} {Phys. Rev. D}\ }\textbf {\bibinfo {volume}
  {72}},\ \bibinfo {pages} {084011} (\bibinfo {year} {2005})},\ \Eprint
  {http://arxiv.org/abs/hep-th/0504067} {arXiv:hep-th/0504067} \BibitemShut
  {NoStop}%
\bibitem [{\citenamefont {Ohashi}\ \emph {et~al.}(2015)\citenamefont {Ohashi},
  \citenamefont {Tanahashi}, \citenamefont {Kobayashi},\ and\ \citenamefont
  {Yamaguchi}}]{Ohashi:2015fma}%
  \BibitemOpen
  \bibfield  {author} {\bibinfo {author} {\bibfnamefont {S.}~\bibnamefont
  {Ohashi}}, \bibinfo {author} {\bibfnamefont {N.}~\bibnamefont {Tanahashi}},
  \bibinfo {author} {\bibfnamefont {T.}~\bibnamefont {Kobayashi}}, \ and\
  \bibinfo {author} {\bibfnamefont {M.}~\bibnamefont {Yamaguchi}},\ }\href
  {\doibase 10.1007/JHEP07(2015)008} {\bibfield  {journal} {\bibinfo  {journal}
  {JHEP}\ }\textbf {\bibinfo {volume} {07}},\ \bibinfo {pages} {008} (\bibinfo
  {year} {2015})},\ \Eprint {http://arxiv.org/abs/1505.06029} {arXiv:1505.06029
  [gr-qc]} \BibitemShut {NoStop}%
\bibitem [{\citenamefont {Nutma}(2014)}]{Nutma_2014}%
  \BibitemOpen
  \bibfield  {author} {\bibinfo {author} {\bibfnamefont {T.}~\bibnamefont
  {Nutma}},\ }\href {\doibase 10.1016/j.cpc.2014.02.006} {\bibfield  {journal}
  {\bibinfo  {journal} {Comput. Phys. Commun.}\ }\textbf {\bibinfo {volume}
  {185}},\ \bibinfo {pages} {1719} (\bibinfo {year} {2014})},\ \Eprint
  {http://arxiv.org/abs/1308.3493} {arXiv:1308.3493 [cs.SC]} \BibitemShut
  {NoStop}%
\end{thebibliography}%

\end{document}